\documentclass[a4paper,11pt,reqno,superscriptaddress,nofootinbib, showkeys, aps, pre]{revtex4-2}
\usepackage[centertags]{amsmath}
\usepackage{amsfonts}
\usepackage{amssymb}
\usepackage{amsthm} 
\newtheorem{example}{Example} 
\usepackage{titlesec}
\usepackage{newlfont}
\usepackage{stmaryrd}
\usepackage{mathrsfs}
\usepackage{mathtools}
\usepackage{euscript}
\usepackage{graphicx}
\usepackage{enumerate}
\usepackage{todonotes}
\usepackage{color}
\usepackage{orcidlink}
\usepackage{subcaption}
\usepackage{booktabs}
\usetikzlibrary{decorations.pathmorphing,arrows.meta}
\usepackage{tikz}
\usepackage{pgf}
\usetikzlibrary{positioning,fit,calc}
\usetikzlibrary{arrows,automata}
\usepackage{wrapfig}
\usepackage{amscd}
\usepackage{cancel}
\usepackage{hhline}
\usepackage{import}
\usepackage{tikz}
\usetikzlibrary{arrows.meta}
\usepackage{enumitem} 
\newlist{condenum}{enumerate}{1} 
\setlist[condenum]{label=\bfseries Condition \arabic*., 
                   ref=\arabic*, wide}
 
\usepackage{changes}

\let\oldsqrt\sqrt
\def\sqrt{\mathpalette\DHLhksqrt}
\def\DHLhksqrt#1#2{
	\setbox0=\hbox{$#1\oldsqrt{#2\,}$}\dimen0=\ht0
	\advance\dimen0-0.2\ht0
	\setbox2=\hbox{\vrule height\ht0 depth -\dimen0}
	{\box0\lower0.4pt\box2}}
\newcommand{\id}{\textrm{d}}

\usepackage{url}
\usepackage{lipsum}
\usepackage{mathtools}
\usepackage{lmodern}
\usepackage{anyfontsize}
\usepackage{hyperref}
\newcommand{\tb}{\textcolor{blue}}

\begin{document}
\title{Entropy and Frenesy:\\
the arrow and the bustle of time}

\author{Christian Maes \orcidlink{0000-0002-0188-697X}\\
{\it Department of Physics and Astronomy, KU Leuven}}


\begin{abstract}
Nonequilibrium statistical mechanics is the science of time as it manifests itself in the transient or steady entropy producing dynamical condition of many-body systems. Beyond the vivid multiformity of the phenomena themselves, three foundational issues lie at its heart. The first question concerns the very origin of entropy production, which gives macroscopic processes their direction and thereby defines the arrow of time: how does an atomistic mechanical world give rise to dissipated heat, thermalization, and time-reversal symmetry breaking? The second asks why, far beyond the familiar behavior of equilibrium systems, driving or agitation may produce and select stable and functionally rich structures: what is it about far-from-equilibrium, low-entropy driving that unleashes the wealth and intricacy of the natural world? The third issue is partly methodological: to determine how, and for which subclass of such wildly diverse nonequilibrium systems, a systematic, unified description and modeling from first principles remains possible.\\
This essay addresses these issues in an informal, semi-popular-science style, offering a personally engaging yet wide-ranging introduction to the complementary roles of entropy production and the less familiar, more recently coined notion of frenesy in the dynamical fluctuations of steadily driven systems. Together, we argue, they provide a more complete starting point for understanding physical time on mesoscopic to macroscopic scales.
\end{abstract}
\;\qquad


\maketitle
\newpage
\tableofcontents
\newpage
\section{Introduction}
\begin{figure}[htbp]
\centering
\begin{tikzpicture}[
    >=Stealth,
    every node/.style={font=\normalsize}
]

\draw (-6.5,-2.5) rectangle (6.5,2.5);

\node[align=center] at (-4.3,0) {Dynamical\\ fluctuations};

\node[align=left] at (3.3,0) {Response \\ Relaxation};
\node[align=left] at (4.7,0) {relations};

\draw[double distance=2pt, -{Stealth[length=4mm]}] (-1.8,0) -- (1.8,0);

\node[align=center] at (0,2) {dissipation excess};
\draw[->, thick] (0,1.4) -- (0,0.3);

\node[align=center] at (0,-2) {frenesy excess};
\draw[->, thick] (0,-1.6) -- (0,-0.3);

\end{tikzpicture}
\captionsetup{justification=raggedright, singlelinecheck=false}
\caption{{\small Stimuli give rise to dissipation and frenesy excesses.  The structure of the dynamical fluctuations underlies the corresponding response and relaxation relations. When starting from a nonequilibrium condition, changes in frenesy matter considerably.}}
\label{fig:dynfluct-scheme}
\end{figure}
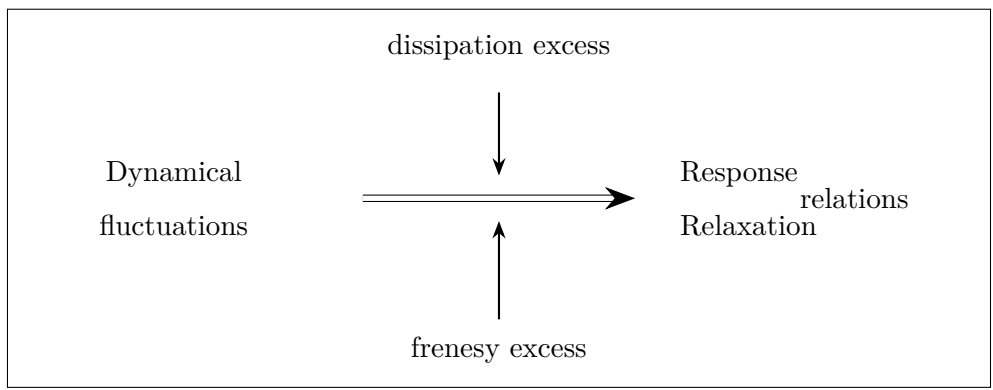

Harmony between seemingly contrasting elements has long inspired play, music, and literature.  Statistical mechanics of open and driven systems has its own unlikely pair: entropy production and frenesy, both defined on the system trajectories, but with the last one still less familiar. One tells us about irreversibility: entropy production is antisymmetric under time-reversal.  The other, frenesy, is about time-symmetric dynamical activity as manifested in waiting times and accessibility. It simply counts how much is happening: how many jumps, switches, or collisions occur, regardless of which direction they go.  As a one-liner: the entropy production of a trajectory tells ``how strong is time's arrow?''; frenesy asks ``how much does it fidget?''\\  In the present essay, part physics review, part cultural history, part manifesto, we emphasize the role of that second element, frenesy, which in the title we call the ``bustle of time'': its liveliness, its fidgeting, independent of any direction.  However, neither alone suffices to tell the full story of time.
Both play their role in estimating the plausibility of system trajectories, and both are best understood as components of dynamical fluctuations.  At that level, time-symmetric and time-antisymmetric fluctuations play complementary roles in determining how systems respond to their environment (see Fig.~\ref{fig:dynfluct-scheme}). Given a stimulus, the response is built from an entropic and a frenetic contribution.

\begin{figure}[htbp]
\centering
\begin{tikzpicture}[
    >=Stealth,
    every node/.style={font=\normalsize}
]

\draw (-6.5,-3.0) rectangle (6.5,2.5);

\node[align=center] at (-4.5,0) {Driving and\\ dissipation};

\node[align=center] at (4.5,0) {Dissipative\\ structure};

\foreach \x in {-2.2,-1.8,...,1.8}{
    \fill[gray!50] (\x,-0.15) rectangle (\x+0.35,0.15);
    \draw[gray!30] (\x,-0.15) rectangle (\x+0.35,0.15);
}
\draw[-{Triangle[length=4mm,width=5mm]}, gray!60, line width=1pt] (2.15,0) -- (2.5,0);

\node[align=center] at (0,2) {frenetic selection};
\draw[->, thick] (0,1.6) -- (0,0.3);

\node[align=center] at (0,-2) {dissipation-induced instability\\\quad \;
driving-induced stability};
\draw[->, thick] (0,-1.4) -- (0,-0.3);

\end{tikzpicture}
\captionsetup{justification=raggedright, singlelinecheck=false}
\caption{\small{Some (even narrow) new dissipation channel can destabilize what was boring behavior, and a large enough driving can stabilize what was forbidden in thermodynamic equilibrium.  In both cases, the way to interesting so-called dissipative structures is paved or mediated by time-symmetric aspects of the kinetics. We argue that frenesy selects among the possible structures a driven system could adopt.}}
\label{fig:frenetic-scheme}
\end{figure}
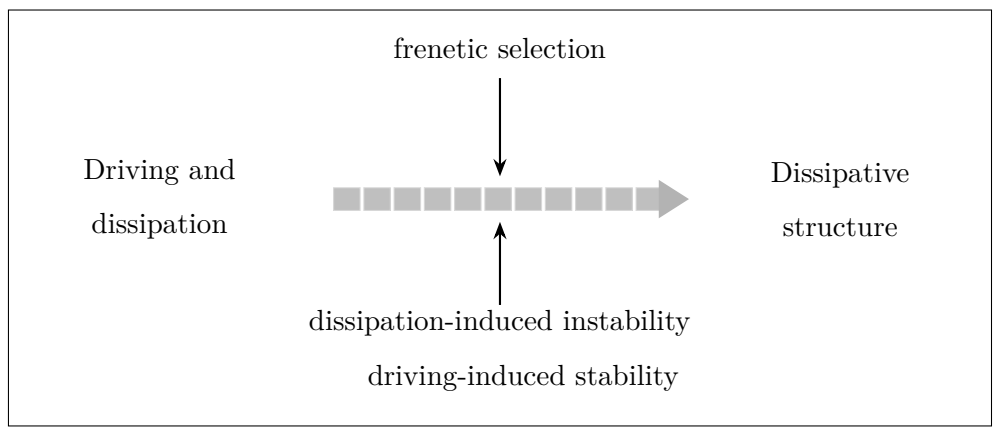

Although the fluctuations of the entropy production, at least for an interesting class of nonequilibrium systems, remain constrained by (a local version of) microscopic reversibility, dissipation in general is the hallmark of time's arrow. For a long time, it has been associated with losses and inefficiency.  Yet, today, entropy production is no longer viewed merely as waste or loss of efficiency. 
Rather, it appears to enable richer and more adaptive behavior as compared to the equilibrium case where there is no net dissipation.  It does so by bringing into play the dependence on waiting times, reactivities, and escape rates along system trajectories, allowing sensitivity and response to external stimuli that differ fundamentally from their equilibrium counterparts.
 We need to explain the role of these time-symmetric aspects of the fluctuating dynamics, measuring activity {\it vs.} quiescence, in paving the route from dissipation to structure (as  in Fig.~\ref{fig:frenetic-scheme}).

After all, the universe harbors life and the relaxation of low-entropy initial conditions has created new phases, unexpected patterns and complicated forms of organization.  But also familiar and less sensational examples abound. Friction allows the tippe top to spontaneously invert itself and spin on its narrow stem\footnote{As first recognized in 1952, the inversion of the tippe top is impossible without sliding friction, as treated in different ways by different authors including Kees Braams, Nico Hugenholtz \cite{Hugenholtz1952} and Jan Haringx. Far from merely removing energy, friction creates the torque that allows the spinning top to rise onto its stem.}. Likewise, birds and airplanes rely on the (even very small) viscosity of air, through the viscous dissipation it entails, to generate the boundary layers, circulation, and vortex wakes that make sustained flight possible\footnote{While the outer flow is largely inviscid, a thin viscous boundary layer smooths the flow leaving the trailing edge.  In that way it selects the physically realized circulation (value of the integral of the flow velocity along a loop enclosing the wing) in mathematical correspondence with the Kutta-Prandtl condition (1902-1916). Along these lines, also the  d'Alembert paradox (1752) finds its solution \cite{Darrigol2005,Grimberg2008}.}.\\

Section \ref{mech} stands at the beginning to remind us of clock-time: microscopic laws are time-reversible.  However, the point remains that also in the context of relativity, time is associated to trajectories: clocks along different trajectories can record different elapsed times if reunited.\\
Sections \ref{cla}, \ref{bol}, \ref{fac} deal with entropy and its production, mostly elementary and well-known to the science and engineering student except for the story about {\it the house with the narrow doorways}.  There and going beyond the pioneering insights of Clausius and Boltzmann, we emphasize that especially far away from equilibrium, the probability of a trajectory on macroscopic scales also depends on the accessibility of the successive macroscopic conditions, not just on their entropy, while size of the phase space volumes of course does matter a great deal for the entropy to increase.\\ 
Section \ref{fr} brings the newest actor on stage.  Frenesy is the time-symmetric part in the fluctuation functional for dynamical fluctuations.  Local detailed balance, known for about 70 years, was the missing ingredient that finally allowed fluctuation theorems and frenesy to be formulated precisely.  The response of a nonequilibrium system depends also on the change in frenesy.  It explains the unrest --- literally, the oscillating {\it Unruh}\footnote{the balance wheel of a clock.} regulated by a clock's escapement --- that dissipation sets free via the so-called frenetic contribution. We add ideas on how to measure that frenesy, and how it fits rather well with recent advances in experimental methods to visualize or unravel particle trajectories.\\
The constructive role of dissipation is highlighted in Section \ref{newe} with newer views on dissipation and driving (by way of a historical detour though, where two Belgians, Nobel prize winners, are prominent). The frenetic component, we argue, is what selects the dissipative structure, illustrated in a simple but musical example by the flute.\\
Section \ref{math} states the resulting thesis and poses some open questions.  Many-body nonequilibrium physics, including collective phenomena in driven and active systems, is a wide-open and fascinating field of study.  Geometric dissipation and quantum trajectories for open systems remain largely unexplored as well.\\

For a more pleasant and selective reading, the essay digresses here and there considerably from its main line. We have refrained from adding more sophisticated examples and illustrations, so as not to slide into technicality too soon.  We have chosen instead a more lyrical, intuitive style, as a way of inviting the reader in.  Focused and more detailed expositions are found in the reviews \cite{nondiss,fren,whatis}.  

\section{Mechanical time}\label{mech}
Time is a concept that structures much of human experience, from daily routines to long-term planning. According to the French pediatrician Françoise Dolto, a child first encounters the passage of time when a parent says ``Wait,'' introducing a symbolic mediation between immediate desire and delayed satisfaction; it helps the child transition from a purely physical existence to becoming a psychological ``subject.''  On the more scientific side, studies in history, the working of memory and consciousness, and the evolution of species and of the universe as a whole, bring time to the foreground. 
Physics in particular has obviously been much occupied with the notion of time.  In classical mechanics, time is absolute and universal, ticking the seconds away in this our universe.  In the Principia, Isaac Newton writes: 
\begin{quote}
Absolute, True, and Mathematical Time, of it self, and from its own nature, flows equably without relation to anything external, and by another name is called Duration.
\\
{\it Beginning of Scholium I in Newton's Philosophi{\ae} Naturalis Principia Mathematica, translation by Andrew Motte, 1729.} 
\end{quote}
Time is then like a medium of succession with events happening one after another.\\

\begin{figure}[htbp]
\centering
\includegraphics[width=0.75\textwidth]{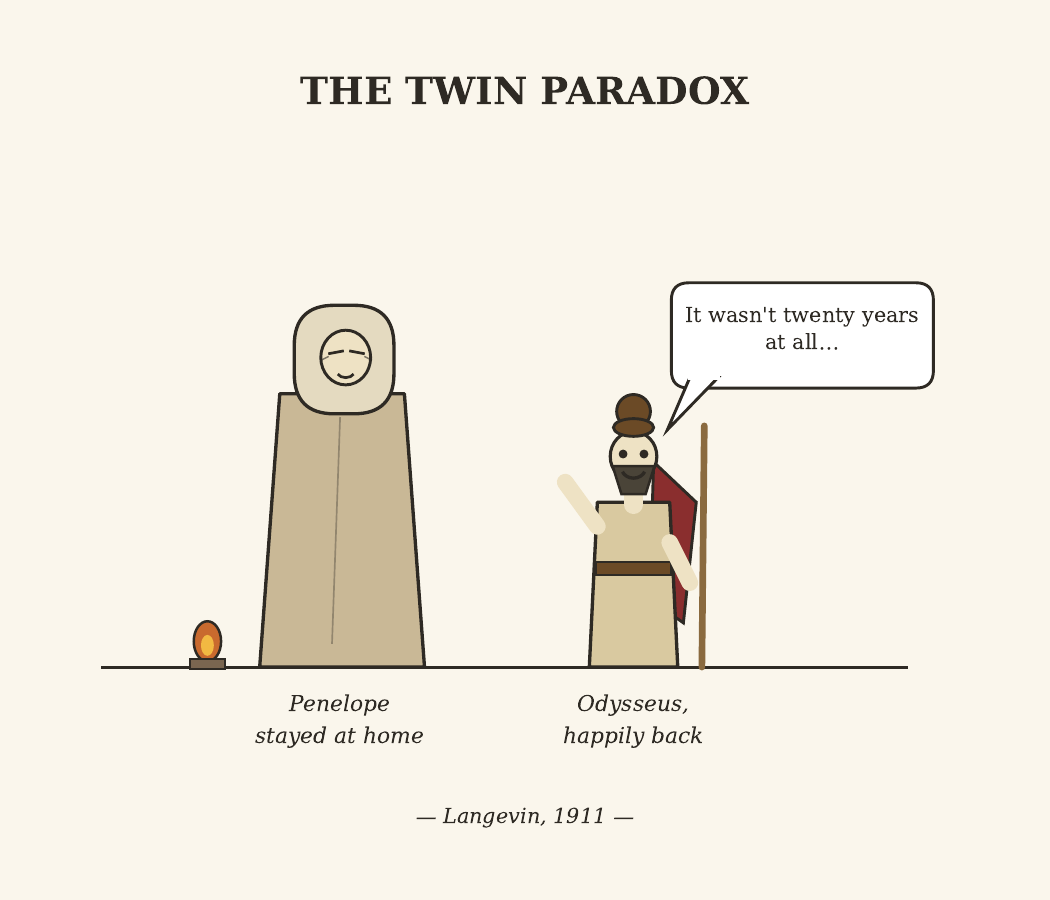}
\captionsetup{justification=raggedright, singlelinecheck=false}
\caption{\small{The Twin Paradox without twins, mythologically recast as a fast Odysseus reaching home.}}
\label{twin_paradox}
\end{figure}

More than 200 years had to pass for a totally different and revolutionary new version of (mechanical) time, which, nevertheless, ``sprung from the soil of
experimental physics'' as Hermann Minkowski was emphasizing in his ``Space and Time'' lecture of 1908. In the words of Erwin Schrödinger,
\begin{quote}
 The special theory of relativity meant the dethronement of time as a rigid tyrant imposed on
us from outside, a liberation from the unbreakable rule of ``before and after.'' ...
it seems to encourage the thought that the whole ``timetable'' is probably
not quite as serious as it appears at first sight. 
\\
--- {\it Tarner Lecture, 1956, published in ``Mind and Matter''}.   
\end{quote}
This revolution, driven largely by Albert Einstein, distinguished between coordinate time, which depends on the chosen reference frame, and proper time, the time actually recorded by a clock along a particular path through spacetime. While coordinate times assigned to events may differ from one observer to another, the proper time along a given worldline is the same for all observers.
Clocks that wander
off along different trajectories through spacetime can have recorded quite different
elapsed times if they ever reunite, as paradoxically illustrated in the famous Twin Paradox, first conceived by Paul Langevin in 1911; Fig.~\ref{twin_paradox}.  That ``paradox'' is just one example of 
the many intricacies of spacetime.\\

Apart from the trajectory-dependence of (even) mechanical time, a second thing to mention
is that basic physics 
is governed by microscopically time-reversal invariant laws in which yesterday and tomorrow play symmetric roles\footnote{We disregard here as irrelevant for the discussion the time-reversal symmetry breaking in the weak interaction.  Knowing that a kaon decay slightly favors one direction does not explain why broken cups do not fix themselves or why heat flows from hot to cold.}. As a simple example within Newtonian mechanics, the position $x$ and velocity $v$ of a point particle of mass $m$ in one spatial dimension are related via evolution equations of the form
\[
\frac{\id}{\id t} x(t) = v(t),\qquad m\,\frac{\id}{\id t} v(t) = F(x(t))
\]
where $F$ denotes the time-independent force.
Clearly, flipping the velocity $v\rightarrow -v$ while changing the arrow of time $t\rightarrow -t$ leaves the equations of motion invariant and the time-reversed motion is subject to the same forces.  Here is how James Clerk Maxwell describes the time-reversal invariance of Newtonian mechanics:
\begin{quote}
    Now one thing in which the materialist (fortified with dynamical knowledge) believes is that if every motion, great and small, were accurately reversed, and the world left to itself again, everything would happen backwards; the fresh water would collect out of the sea and run up the rivers, and finally fly up to the clouds in drops, which would extract heat from the air and evaporate, and afterwards in condensing would shoot out rays of light to the sun, and so on. Of course all living things would retrograde from the grave to the cradle, and we should have a memory of the future and not of the past.\\
    {\it From Maxwell's private correspondence  and later reused nearly verbatim in the anonymous 1878 review of Tait’s book ``Thermodynamics'' (published in Nature).}
\end{quote}
It can be read as a prescient commentary on the issues that Ludwig Boltzmann was grappling with (see below in Section \ref{bol}).
Einstein was strongly influenced by Boltzmann as well, and repeatedly emphasized that the fundamental equations of physics, including relativity, do not contain an intrinsic distinction between past and future.  A reversed-entropy universe would contain observers whose memory records point toward what we now call the future, and our psychological experience of temporal passage is not built into the microscopic equations of motion.  That view also survived the quantum revolution, even though time and motion through space are traditionally (and unfortunately) less explicitly dealt with there. \\ 

Looking at trajectories and their temporal behavior reveals two further faces of time. The first is its unmistakable direction in macroscopic processes, the arrow of time, governed by the relentless increase of entropy. The second is its restless pulse as experienced by open systems in contact with media and reservoirs, where the ceaseless activity and traffic of particles are summarized by the notion of ``frenesy.''  
It is to these aspects of time, 
entropic and frenetic, that we turn in the following sections.

\section{Clausius entropy}\label{cla}
Maxwell defined thermodynamics as the investigation of the ``relations between the thermal and the mechanical properties of substances''  (1871). Einstein  in his ``Autobiographical Notes'' of 1949 called thermodynamics a superior field of science, being ``the only physical theory of universal content of which I am convinced that, within the framework of the applicability of its
basic concepts, it will never be overthrown.''  Cornerstones are the First and the Second Law.  The First Law of thermodynamics has a direct analogue in classical mechanics where it begins with the work-energy theorem: in simple phrasing, the change in kinetic energy equals the work done on the system.  Its thermodynamic formulation plays on a macroscopic scale and involves the somewhat enigmatic notion of heat.  It enters the First Law to mediate between the change in macroscopic energy and the work done on the system.  Heat is energy transfer between a system and its environment due to a temperature difference, not associated with mechanical work\footnote{
Work represents an ordered, macroscopic transfer of energy driven by or yielding a Newtonian force. When a moving fluid stream hits a turbine blade, it does work. If it simply mixes into a region of lower temperature, the energy exchange driven by that temperature gradient is accounted for as heat transfer.}.  
It is not a state function, at least not always, and we often denote it, following Rudolf Clausius, as $\delta Q$ for an elementary transfer.\\
\begin{figure}[htbp]
\centering
\includegraphics[width=0.95\textwidth]{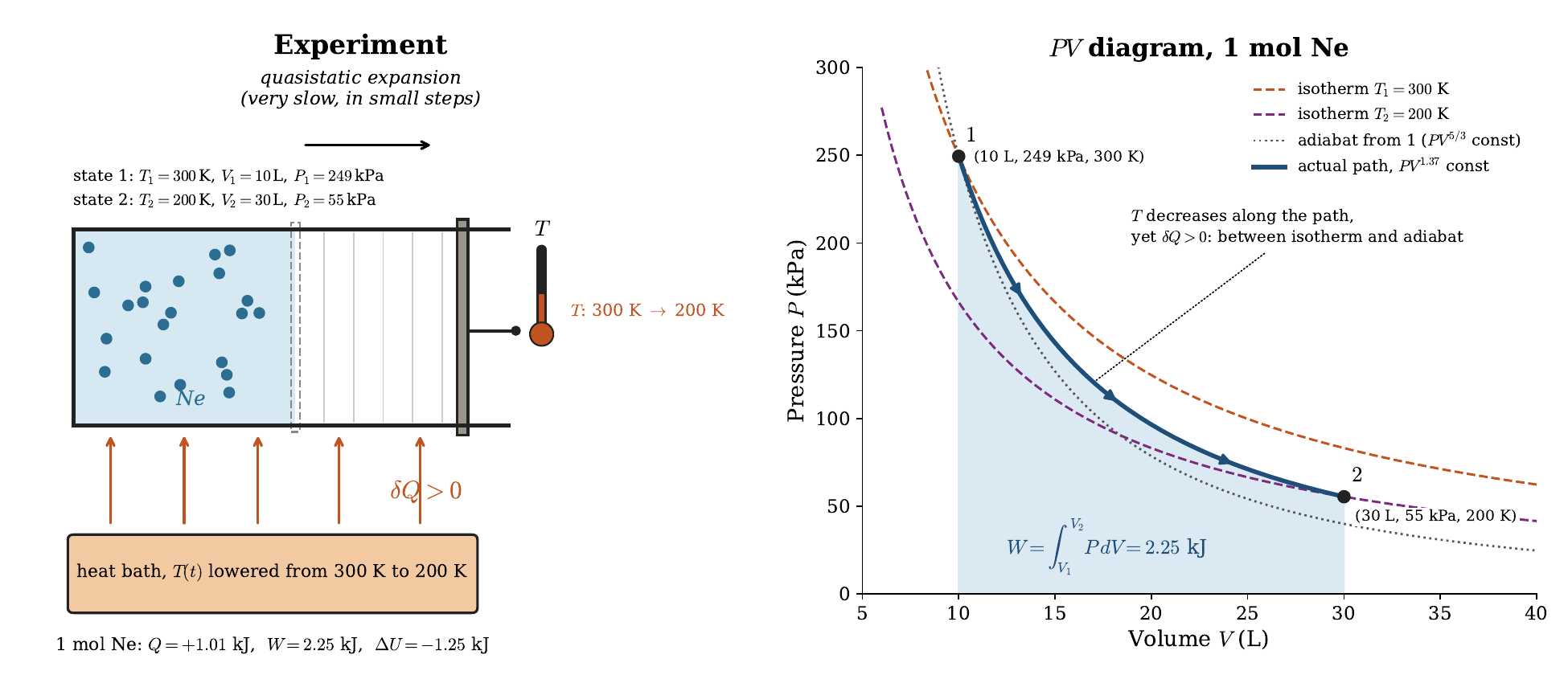}
\captionsetup{justification=raggedright, singlelinecheck=false}
\caption{\small{Quasistatic expansion of one mole of Neon gas along the polytrope $PV^{n}=$ const, $n=1+\log(3/2)/\log 3\simeq1.37$, from $(V_1,T_1,P_1)=(10\,\text{L},300\,\text{K},249\,\text{kPa})$ to $(V_2,T_2,P_2)=(30\,\text{L},200\,\text{K},55\,\text{kPa})$. The path lies between the isotherm and the adiabat through state 1: the gas absorbs heat, $Q=+1.01$\,kJ, but cools, because it does more work, $W=2.25$\,kJ, than the heat supplied, so that $\Delta U=\tfrac32R\,\Delta T=-1.25$\,kJ. Clausius' heat theorem is verified since $\frac 1{T}\delta Q^\text{rev} = R\,\id (\,\log (T^{3/2} \, V)\,)$: both sides give $\Delta S=R\log\big[(T_2/T_1)^{3/2}V_2/V_1\big]=4.08$\,J/K.}}
\label{qsstatic}
\end{figure}

The heat theorem (1865) by  Clausius is based on the idea of a quasistatic process.  For example, a process in which the change of the volume of the system is very slow, or when changes in the temperature of the environment are slow compared to relaxation times of the system.
At each time the system is in equilibrium for the present parameter values.   Those are the reversible evolutions of thermodynamics that are discussed as curves in $PV-$diagrams etc.; see Fig.~\ref{qsstatic} for an ideal gas.  We write $\delta Q^\text{rev}$ for the elementary heat in such a reversible process.  Clausius found that if we divide the reversible heat to the system by the instantaneous absolute temperature $T$,
\begin{equation}\label{ckl}
\frac 1{T}\delta Q^\text{rev} = \id S
\end{equation}
we find a total differential: the `$\id$' indicates the change in a state function, here the thermodynamic entropy $S$ of the equilibrium system.  That is the Clausius heat theorem
and it defines the entropy  as used in the thermodynamics of equilibrium systems and their quasistatic transformations. Clausius chose the name `entropy' because it sounds like energy and it could be adopted in many modern languages without translation. 
In his landmark 1865 paper ``The Mechanical Theory of Heat''  Clausius explains that he coined entropy from the ancient Greek word `trope', meaning ``transformation'' or ``a turning.''  Initially, Clausius referred to this quantity as the ``transformation content'' of a body, as the measure of how energy changes and disperses within a system. \\
\\
The entropy of the system is $S$ and the total change of entropy in the world (system plus environment at temperature $T$) for general processes is 
\[
\id S_\text{total} = \id S - \frac 1{T}\delta Q \geq 0
\]
When Clausius analyzed the Carnot heat engine he found that it was the increase $\Delta S_\text{total} > 0$ that is directly responsible for the dissipation and non-optimal efficiency.  The Second Law states that the entropy of a closed mechanical macroscopic system increases to become constant when reaching thermodynamic equilibrium.  It translates our experience with macroscopic processes where the energy becomes less available for doing useful work --- the arrow of time. It is a reality that demands our attention. Here is the wisdom of Clausius as admonition to our generation:
\begin{quote}
    The following centuries will have the task of introducing a wise economy in the use of what the sources of power in nature offer to us, and especially of not squandering wastefully what we find in the soil as a legacy of earlier times, and what cannot be replaced by anything.\\
{\it R. Clausius, \"Uber die Energievorr\"athe der Natur und ihre Verwerthung zum Nutzen der Menschheit (``On the Energy Reserves of Nature and Their Utilization for the Benefit of Humanity''), 1885}.
\end{quote}
In other words, progress requires not only power, but restraint --- a Confucian sentiment, as when Mengzi (Mencius), about twenty-two centuries earlier, is concerned with (also) renewable resources:
\begin{quote}
  If the fine nets are not allowed into the large ponds, the fish and turtles will be more than can be consumed. If axes and bills enter the hills and forests only in the proper seasons, the timber will be more than can be used.\\
  {\it  Mengzi in Book I, Part A.}
\end{quote}

\section{Boltzmann entropy}\label{bol}
We mentioned that the First Law of thermodynamics has an analogue in classical mechanics.  The natural follow-up is clarifying how the Second Law emerges from mechanics: is there a microscopic derivation of the increase of thermodynamic entropy for a closed and isolated macroscopic system?\\
That question has stirred many minds as we, unifiers and reductionists, face a problem: the more fundamental microscopic laws are time-reversible, while macroscopic behavior shows time-asymmetry.  How can these facts be reconciled?  The origin of dissipation starting from time-reversible laws following Hamiltonian or unitary dynamics continues to confuse\footnote{the information paradox being the newest.}.  Here is the advice of Schr\"odinger:
\begin{quote}
The spontaneous transition from order to disorder is the quintessence
of Boltzmann’s theory... This theory really grants an understanding
and does not reason away the dissymmetry of things by means of an
a priori sense of direction of time... No one who has once understood
Boltzmann’s theory will ever again have recourse to such expedients. It
would be a scientific regression beside which a repudiation of Copernicus
in favor of Ptolemy would seem trifling.\\
--- {\it In ``Statistical Thermodynamics,'' 1946}. 
\end{quote}
The bridge between the micro-world and macro-behavior is entropy.
Entropy is not about messiness. A shuffled deck of cards is not high entropy because it looks disordered to humans. Boltzmann instructed us to count and estimate plausibility via statistical methods.  He went beyond Clausius by assigning entropy to individual macroscopic nonequilibrium conditions.  Entropy is then about counting the microscopic possibilities to realize a macroscopic condition defined by a physically relevant choice of macroscopic variables.  Boltzmann’s famous idea can be summarized in a single equation:
\begin{equation}\label{si}
S = k_\text{B}\, \log W
\end{equation}
Here, $W$ represents the number of microscopic arrangements (in positions and momenta of all particles) compatible with the macroscopic condition for which we seek the entropy (as defined via the macroscopic variables), and with $k_\text{B}$ Boltzmann's constant (as introduced by Max Planck). The larger the number of hidden possibilities, the larger the entropy. A glass of water with all its molecules evenly mixed can be realized in unimaginably many microscopic ways. A glass where all fast molecules sit on the left and all slow molecules on the right can be realized in, comparatively, very few ways. Following Laplace's indifference (for the microcanonical ensemble) it seems natural to estimate plausibility in terms of that entropy: a mixed glass is more likely, not because nature ``prefers disorder,'' but because there are overwhelmingly more ways to be mixed than unmixed.\\

In classical mechanics, counting amounts to evaluating the phase-space volume on a fixed-energy surface $\Omega$\footnote{Yet, in Boltzmann's work, atomism played a crucial role and even phase space got discretized, as we would do today via Planck's constant $h$.}.
The Boltzmann entropy of a micro-state $X$ (the positions and momenta of a huge number of particles in a fixed volume at fixed energy) is
\begin{equation}\label{sil}
S(X) = k_\text{B} \,\log W(M(X))
\end{equation}
where $W(M(X))$ is the Liouville volume of the macro-state $M(X)$ to which $X$ belongs, defined on the energy shell $\Omega$ of a closed and isolated mechanical system.  It is important to repeat that the map $X\to M(X)$ is many-to-one.  That is the physical coarse-graining to thermodynamic or hydrodynamic variables, \cite{sasa2014}.\\

From formula \eqref{sil}, Boltzmann argued how that entropy {\it typically} increases along microscopic trajectories $X(t)$ of a macroscopic system when relaxing to equilibrium. Also the formation by gravity of stars and galaxies, the clustering of mass, is increasing the entropy because gravitational potential energy is converted into kinetic energy, enormously increasing the accessible phase-space volume of the microscopic degrees of freedom (the phase space volume associated to moving at larger velocities)\footnote{The Jeans instability is the onset of that gravitational collapse, explaining why clustering starts in Newton's theory of gravitation.}.  
Indeed, it is fair to expect that the system's (microscopic) trajectory typically goes through macroscopic regions $M(X(t))$ of increasing entropy as long as entropy is assumed to measure plausibility. We can then sometimes prove, in the limit where the macroscopic variables are autonomous ({\it i.e.}, macroscopically reproducible under the dynamics) that $S(X(t))$ is increasing in time $t$ for the mechanical evolution; a very simple argument is included in \cite{dmn2003}.
Asymptotically, when the equilibrium condition is reached, the limiting entropy is the Clausius thermodynamic entropy, which is the Boltzmann entropy at thermodynamic equilibrium. It does not mean that the reverse of what is the typical macroscopic evolution (where the entropy increases by $\Delta S = S_\text{high} - S_\text{low}$) is impossible. Rather, the probability that a system found in the high-entropy macrostate is later observed in the low-entropy one is low by an amount 
\begin{equation}\label{dar}
\text{Prob}[S_\text{high} \rightarrow S_\text{low}] \propto \exp[-\Delta S/k_B]
\end{equation}
For the free expansion of 1 mole of an ideal gas to double its volume, $\Delta S = 5.763$ J/K and $\Delta S/k_B \simeq 4.17\times 10^{23}$.
That shows how dissipation quantifies the pointed arrow of time.\\
\begin{figure}[htbp]
\centering
\includegraphics[width=0.7\textwidth]{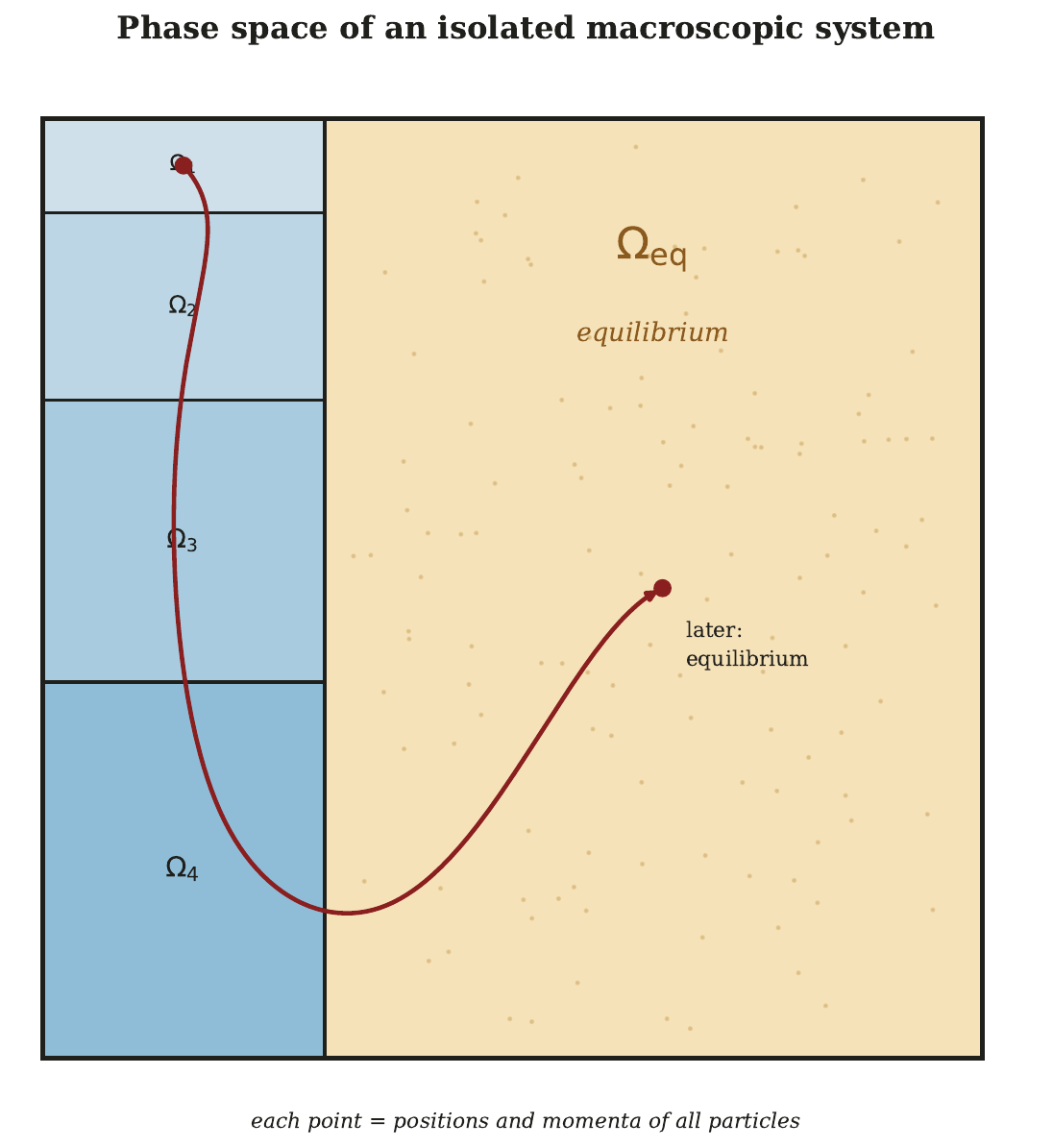}
\captionsetup{justification=raggedright, singlelinecheck=false}
\caption{Phase space of an isolated macroscopic system, partitioned into macrostates. A single microscopic trajectory, starting in a small-volume macrostate, wanders through rooms of increasing phase-space volume before reaching the overwhelmingly large equilibrium macrostate.}
\label{fig:phase-space-rooms}
\end{figure}

That insight has been reformulated in many ways,  stressing the point that the Second Law is truly a statistical law --- a {\it moral certainty}, as Maxwell described it in  a letter to John William Strutt (Lord Rayleigh) in 1870. Also Josiah Willard Gibbs acknowledged the probabilistic nature of the Second Law, concluding that a spontaneous decrease of entropy on a macroscopic scale is not strictly impossible, but rather vastly improbable based on the principles of phase space and Liouville's Theorem. He famously noted that
\begin{quote}
    the impossibility of an uncompensated decrease in entropy seems to be reduced to  improbability.\\
    {\it in  Gibbs' foundational 1875 paper, ``On the Equilibrium of Heterogeneous Substances''}
    \end{quote}  
    Similarly, in his {\it Introduction to Statistical Physics}, the physicist Wang Zhuxi engaged with the deep questions of time's asymmetry  and the microscopic basis of irreversibility. He taught  
    that the macroscopic progression of time, where entropy always increases in an isolated system, is a statistical effect:
\begin{quote}
The macroscopic arrow of time is a statistical manifestation. While the microscopic equations of motion are symmetric with respect to time reversal (\(t \to -t\)), the overwhelming probability of finding the system in a state of maximum disorder dictates that entropy increases, establishing an irreversible directionality for time.\\
--- {\it Wang Zhuxi}
\end{quote}
Obviously, these arguments distinguish past and future only to the extent that we assume ``special'' initial conditions; they deal with the transition from `order' to `disorder.' We refer to  more detailed expositions, including \cite{lebowitz_goldstein2004boltzmann,lebowitz2008timesarrow}, graceful but subtle, also to counter older and more recent criticism\footnote{Older paradoxes go under the name of Zermelo (Poincaré recurrence seems to forbid permanent entropy increase) and of Loschmidt (how can reversible microscopic laws yield irreversible macroscopic behavior).  More recently, the transition from microscopic reversibility to macroscopic irreversibility is sometimes mischaracterized in terms like `incomplete information' or `imperfect boundary conditions' and the role of coarse-graining gets detached from its meaning in thermodynamics.  We may also add that the Boltzmann analysis refers in the first place to local arrows of time, with no strict need to assemble those into one single, universe-wide direction of time.}. \\ 

Boltzmann's picture of phase space provides a remarkably powerful explanation of the Second Law of thermodynamics, and even more: it tells us that a system almost inevitably evolves toward equilibrium because equilibrium corresponds to an enormously larger region of phase space than any nonequilibrium condition. In a sense, equilibrium is simply where there are overwhelmingly many more ways for the system to be.
Yet this elegant argument leaves an important question unanswered: how long does it take to get there? When a constraint is removed, why does the system not reach equilibrium immediately? Why do some systems relax in a fraction of a second, while others take hours, years, or even longer?\\
Boltzmann's argument is essentially a static one. It compares the sizes of different regions (macroscopic values) in phase space, treating larger regions as more probable irrespective of their accessibility. But the speed of relaxation depends not only on the size of the destination, but also on how easily it can be reached.\\
Imagine a large house. The biggest room may be the natural place to store all the furniture, but moving everything there depends on much more than the room's size. Narrow doorways, winding corridors, and steep staircases can make the move slow and difficult. Likewise, in phase space, the ``routes'' connecting different regions matter as much as the regions themselves.
A card game offers another analogy. If every card slides freely, a few shuffles quickly produce a random deck. But if the cards stick together, ordinary shuffling no longer mixes them efficiently. The deck may remain far from random for a long time.\\
Many physical systems behave in a similar way. Their dynamics can be hindered by kinetic constraints that make certain microscopic rearrangements extremely difficult. Traffic-like jamming in granular matter and glasses are familiar examples. In such systems, reaching equilibrium is not prevented because equilibrium is unlikely, but because the available pathways are narrow or blocked.  That also forms the dynamical underpinning of metastability where nucleation opens the door to equilibration.
Boltzmann's phase-space argument therefore explains why equilibrium is the overwhelmingly likely destination, but not how the journey unfolds. To understand the speed of relaxation, and especially the behavior of systems far from equilibrium, we must look beyond the volume of phase-space regions and consider the accessibility of those regions and the likelihood of the paths connecting them.\\
  Such considerations are taken up in Section \ref{fr} about frenesy. They reveal the obvious fact that the plausibility of trajectories and the speed of relaxation depend on escape rates and reactivities as well.  Those can be expected more relevant when staying away from equilibrium and are time-symmetric, referring to the openness of the transition channel, like a door between two rooms.  In summary, entropy gives the volume of possibilities; frenesy gives the vigor with which a dynamical system may explore them.  Boltzmann's entropy measures the volume of macrostates in phase space. Frenesy characterizes the dynamical traffic between those regions.  Put differently, the entropy $S$ weighs each destination; frenesy governs journeys.

\section{The many faces of entropy}\label{fac}
Understanding entropy is one of modern science’s greatest triumphs. It helps to translate the microscopic movements of individual particles into macroscopic rules of nature. Far from being merely a measure of disorder, entropy drives spontaneous movements (via entropic forces) and connects the random fluctuations of a system to its overall energy dissipation.  In that sense, the concept of entropy\footnote{We often mix the concepts of entropy, dissipation, entropy flux, and entropy production. Unlike the entropy $S$
discussed above, entropy flux and entropy production live most happily on trajectories, not states.} is protean: it has many faces. Different definitions, from Clausius entropy which is related to heat, over $H$-functionals for gradient dynamics, to Boltzmann's statistical, Gibbs' thermodynamic and Shannon's coding perspectives, all appear to coincide (and can be called `entropy').  That remarkable coincidence is one of the main achievements and simplifications in equilibrium statistical mechanics. Heat and motion got connected with combinatorics, generating the arrow of time, and producing thermodynamic forces, while making the near-equilibrium condition understandable from energy-entropy considerations.  Equilibrium is exceptional precisely because these different notions collapse into one another.  It is powerful and the influence of entropy extends well beyond thermodynamics. For instance, variational principles inspired by statistical physics now underpin algorithms for optimization, associative memory, and machine learning. In that sense it is fitting that the Nobel Prize Physics 2024 (awarded to John Hopfield and Geoffrey Hinton) acknowledged utilizing tools from statistical physics to build the foundational architectures of modern artificial intelligence. 
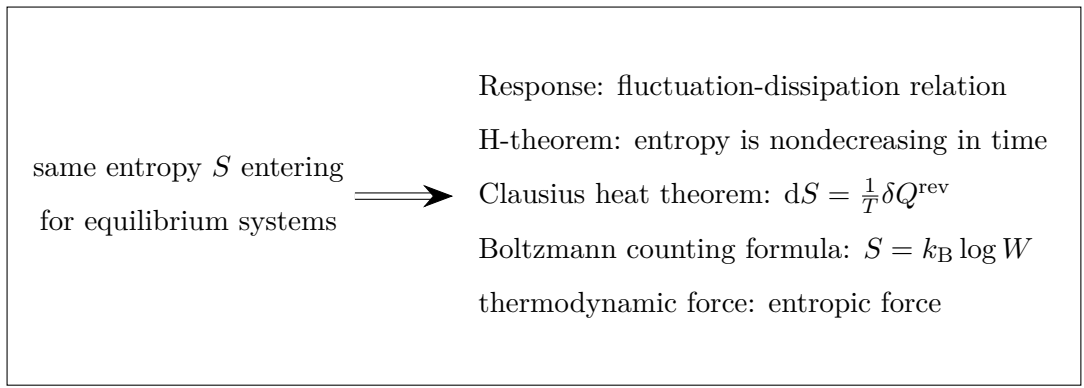
\begin{figure}[htbp]
\centering
\begin{tikzpicture}[
    >=Stealth,
    every node/.style={font=\normalsize}
]

\draw (-6.7,-2.5) rectangle (7.5,2.5);

\node[align=center] at (-4.3,0)  {same entropy $S$ entering \\ for equilibrium systems};

\node[align=left] at (3.3,0) {Response: fluctuation-dissipation relation \\ H-theorem: entropy is nondecreasing in time\\ Clausius heat theorem: $\id S = \frac 1{T}\delta Q^\text{rev}$ \\ Boltzmann counting formula: $S= k_\text{B}\log W$ \\ thermodynamic force: entropic force};
\draw[double distance=2pt, -{Stealth[length=4mm]}] (-2.1,0) -- (-0.8,0);
\end{tikzpicture}
\captionsetup{justification=raggedright, singlelinecheck=false}
\caption{\small{Many faces showing Clausius and Boltzmann formul{\ae}, the monotonicity of the entropy, its appearance in the fluctuation-dissipation relation and its role as statistical force.}}
\label{fig:equilibrium-tshirt}
\end{figure}

However, when we look at systems that are pushed further away from a balanced equilibrium condition, such as driven systems, these neat connections fall apart. Equilibrium creates the illusion that all these faces are one and the same. Nonequilibrium reveals that they are different.  Clausius entropy and Boltzmann entropy are no longer equal \cite{negcap}. The dynamics itself becomes an essential player, and we see the differences between the various faces of entropy.  Natural intelligence and biological functioning, transport phenomena, macroscopic chaos, and matter under extreme conditions all belong to nonequilibrium statistical mechanics. In that setting, the time-symmetric (or, frenetic) fluctuation sector and its associated kinetics separate concepts that coincide at equilibrium, distinguishing the various notions of entropy that equilibrium tends to merge.\\

Then, a central challenge for modern physics is to develop a unified framework that explains how the diverse kinetic and statistical aspects of nonequilibrium phenomena fit together within a single coherent theory.  To bridge the gap between simple equilibrium and complex natural systems, we are seeking a powerful methodology explaining in a unified way some of the nonequilibrium phenomena for at least some class of systems. That class of systems, we suggest, contain those that are weakly coupled to well-separated thermal reservoirs.  More specifically, we use models verifying local detailed balance \cite{ldb}.  There, the probability of trajectories is naturally formulated in terms of ensembles, whose action functionals admit direct physical interpretations; see next Section \ref{fr}: the trick is to split the path probability into a part that flips sign under time-reversal (entropy production) and a part that does not (frenesy).

\section{Frenesy}\label{fr}

Over the last decades, physicists studying systems far from equilibrium discovered something surprising. Examples were found for the motion of molecular motors, chemical reaction networks and enzyme kinetic proofreading, including  \cite{pei2026transferActiveMotion,Beyen2026Rayleigh,astumian2024kinetic,astumian1997,astumian2012,Hopfield1974,astumian2019, penocchio2019, rao2016,Penocchio2024Multicycle,Maes2015}. Two systems may be driven in exactly the same way and produce the same entropy, yet take different paths and select different stationary populations and currents. One
dynamics appears calm and sluggish, another restless and eager to change. Entropy production alone cannot distinguish between them, \cite{nondiss}.\\
To name this hidden aspect of dynamics another concept, frenesy, was introduced, and we speak more generally about the frenetic contribution and about frenetic aspects when time-symmetric kinetics is fundamentally involved.\\

The origin of the concept is in the derivation of response relations describing the behavior of nonequilibrium systems subject to some stimulus, basically as illustrated below for the biased random walk. Fluctuation theory for the
entropy production fails to give information about the time-symmetric sector. That was the context in which ‘‘frenetic’’ first appeared in 2006 \cite{MaesVanWieren2006}. Frenesy takes the sound of ‘‘entropy’’  and ‘‘energy’’, but also connects with its meaning as ‘‘frequency’’. Frenesy is directly related to the
words ‘‘frenzy’’ and ‘‘frantic’’, and originates from the Greek $\phi \rho \acute{\eta} \nu$ (phren). In Homer and classical Greek,  $\phi \rho \acute{\eta} \nu$ originally referred to the midriff or diaphragm, but very early it came to denote the seat of thought, feeling, and will.  Etymologically (or poetically), frenesy is thus closer to ``animation'' and liveliness is a good English rendering of just that aspect of time: how the system breathes, independent of the direction of its evolution --- frenesy as the vibrant pulse of time.

\subsection{Response relations}
Understanding response is at the heart of the scientific enterprise. We want to know how a macroscopic system reacts when subject to an external stimulus, even a small one. For microscopic systems, individual responses can be wild: chaos lets tiny causes produce enormous effects. For mesoscopic to macroscopic systems, however, we track smooth changes in statistical averages. 

In the familiar territory of equilibrium, a system's response is governed entirely by the legendary fluctuation-dissipation relation: the way a system reacts to being pushed is perfectly mirrored by how it naturally fluctuates on its own. But far from equilibrium, this simple bridge collapses. To see how frenesy steps in to fix it, we turn to the simplest driven system that exposes the time-symmetric sector: a biased random walk.

\begin{example}[Random walk]\label{rawa}
We take the simplest driven system that already shows how frenesy
enters the response. Picture a dilute suspension of charged colloids pushed
through a toroidal tube of length $L=dn$, filled with a viscous fluid, by a
constant force $\mathcal{E}$ along the tube (Fig.~\ref{fig:toroidal}(a); the charge is absorbed
in $\mathcal{E}$). The colloids are quasi-independent and feel thermal noise,
so one random walker on a ring of $n$ cells, $x=1,\dots,n$ with
$n+1\equiv 1$, stands for all of them (Fig.~\eqref{fig:toroidal}(b)).
\begin{figure}[h]
\centering
    \includegraphics[width=\linewidth]{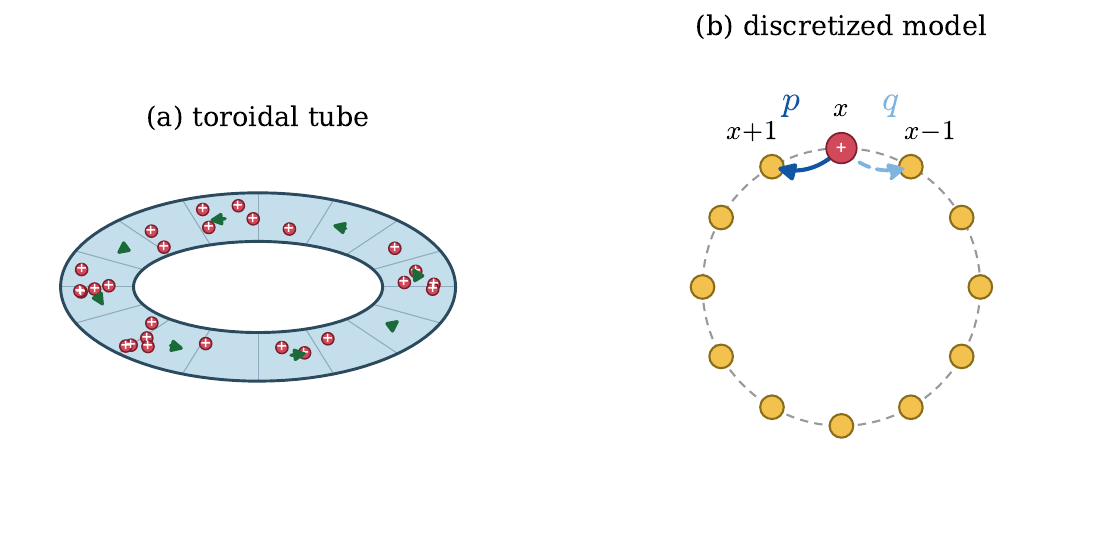}
    \captionsetup{justification=raggedright, singlelinecheck=false}
    \caption{\small{(a) A dilute suspension of electrically charged colloids (red, with net charge $+$) driven through a toroidal tube of length  $L=dn$ filled with a viscous fluid, subject to a constant electromagnetic field $\mathcal{E}$ along the tube (green arrows).
    (b) The corresponding coarse-grained description: a ring of $n$ cells of length $d$, on which the colloid performs a biased continuous-time random walk, hopping to neighboring cells with rates set by the local force $\mathcal{E}$. The $\log$-ratio of the one-step rates is the corresponding entropy flux per $k_\text{B}$ to the thermal bath, $\log p/q = \beta\,\cal E d$, while their sum $p+q$ is the escape rate and may depend on $\cal E$ as well.}}
    \label{fig:toroidal}
\end{figure}

In cell $x$ the walker waits an exponentially distributed time and then jumps
to $x+1$ or to $x-1$.  That is summarized in (transition or hopping rates $p$ (toward $x+1$) and $q$ (toward $x-1$); Fig.~\ref{ju} The total rate of leaving the
cell is the escape rate $\xi=p+q$. Local detailed balance fixes only the ratio
of the two rates: each forward jump transfers the work $\mathcal{E}d$ to the
thermal bath, so
\begin{equation}\label{eq:ldb}
  \log\frac{p}{q}=\beta\mathcal{E}d, \qquad
  p=\frac{\xi\,e^{\beta\mathcal{E}d/2}}{2\cosh(\beta\mathcal{E}d/2)},
  \quad
  q=\frac{\xi\,e^{-\beta\mathcal{E}d/2}}{2\cosh(\beta\mathcal{E}d/2)} 
  \end{equation}
The sum $\xi=\xi(\beta,\mathcal{E})$ is not fixed by local detailed balance
and may depend on the driving; this freedom is exactly where frenesy lives.\\
\begin{figure}[h]
\begin{tikzpicture}[x=0.7cm,y=0.7cm,line cap=round,line join=round,>=stealth, font=\footnotesize]
  \draw[->,thick] (0,-2.4) -- (0,4.7);
  \draw[->,thick] (0,-2.4) -- (13.5,-2.4);
  \node[above] at (0,4.65) { position $x$};
  \node[below right] at (13.5,-2.55) { $\text{time}$};

  \foreach \yy in {-2,-1,0,1,2,3,4} {
    \draw (-0.08,\yy) -- (0.08,\yy);
    \node[left] at (-0.18,\yy) {$\yy$};
  }

  \def\tA{1.5}
  \def\tB{3.8}
  \def\tC{5.0}
  \def\tD{8.0}
  \def\tE{9.9}
  \def\tF{10.7}
  \def\tG{12.2}

  \foreach \tx in {\tA,\tB,\tC,\tD,\tE,\tF,\tG} {
    \draw[densely dotted,gray] (\tx,-2.3) -- (\tx,4.3);
  }

  \node[below] at (0,-2.75) {$0$};
  \node[below] at (\tA,-2.75) {$t_1$};
  \node[below] at (\tB,-2.75) {$t_2$};
  \node[below] at (\tC,-2.75) {$t_3$};
  \node[below] at (\tD,-2.75) { $t_4$};
  \node[below] at (\tE,-2.75) { $t_5$};
  \node[below] at (\tF,-2.75) {$t_6$};
  \node[below] at (\tG,-2.75) { $t_7$};

  \tikzset{closed/.style={circle,fill=blue,inner sep=2.2pt}}
  \tikzset{open/.style={circle,draw=blue,fill=white,line width=0.8pt,inner sep=2.2pt}}

  \draw[line width=1.0pt, blue] (0,0) -- (\tA,0);
  \draw[line width=1.0pt, blue] (\tA,-1) -- (\tB,-1);
  \draw[line width=1.0pt, blue] (\tB,0) -- (\tC,0);
  \draw[line width=1.0pt, blue] (\tC,1) -- (\tD,1);
  \draw[line width=1.0pt, blue] (\tD,2) -- (\tE,2);
  \draw[line width=1.0pt, blue] (\tE,1) -- (\tF,1);
  \draw[line width=1.0pt, blue] (\tF,2) -- (\tG,2);

  \node[open] at (\tA,0) {};
  \node[closed] at (\tA,-1) {};

  \node[open] at (\tB,-1) {};
  \node[closed] at (\tB,0) {};

  \node[open] at (\tC,0) {};
  \node[closed] at (\tC,1) {};

  \node[open] at (\tD,1) {};
  \node[closed] at (\tD,2) {};

  \node[open] at (\tE,2) {};
  \node[closed] at (\tE,1) {};

  \node[open] at (\tF,1) {};
  \node[closed] at (\tF,2) {};

  \node[open] at (\tG,2) {};

  \draw[->] (\tA+0.08,-0.05) .. controls (\tA+0.55,-0.35) and (\tA+0.55,-0.75) .. (\tA+0.20,-0.9);
  \draw[->] (\tB-0.08,-0.95) .. controls (\tB-0.55,-0.75) and (\tB-0.55,-0.25) .. (\tB-0.20,-0.1);
  \draw[->] (\tC-0.08,0.05) .. controls (\tC-0.55,0.30) and (\tC-0.55,0.75) .. (\tC-0.20,0.95);
  \draw[->] (\tD-0.08,1.10) .. controls (\tD-0.50,1.40) and (\tD-0.50,1.85) .. (\tD-0.20,1.98);
  \draw[->] (\tE+0.05,1.95) .. controls (\tE-0.40,1.70) and (\tE-0.40,1.25) .. (\tE-0.14,1.08);
  \draw[->] (\tF+0.04,1.05) .. controls (\tF+0.45,1.25) and (\tF+0.45,1.75) .. (\tF+0.12,1.92);

  \node[above right] at (0.05,0.08) {\scriptsize$x(0)=0$};
  \node[below right] at (\tA+0,-1.05) {{\scriptsize $x(t_1)=-1$}};
  \node[below right] at (\tB-0.2 ,-0.1) {\scriptsize $x(t_2)=0$};
  \node[above right] at (\tC+0.1,0.9) {\scriptsize$x(t_3)=1$};
  \node[above right] at (\tD-0.1,1.9) {\scriptsize $x(t_4)=2$};
  \node[below right] at (\tE-0.45,0.9) {\scriptsize $x(t_5)=1$};
  \node[above right] at (\tF-0.1,2.1) {\scriptsize $x(t_6)=2$};
\end{tikzpicture}
\captionsetup{justification=raggedright, singlelinecheck=false}
\caption{\small{Trajectory of the continuous time random walker of Fig.~\ref{fig:toroidal}(b): waiting separates the jumping times $t_i$. The vertical axis is the position space of possible cell labels $x$. For this trajectory, in \eqref{eq:path}, $N=6, J\,t_7=2 = J\,t_6$.}}
\label{ju}
\end{figure}
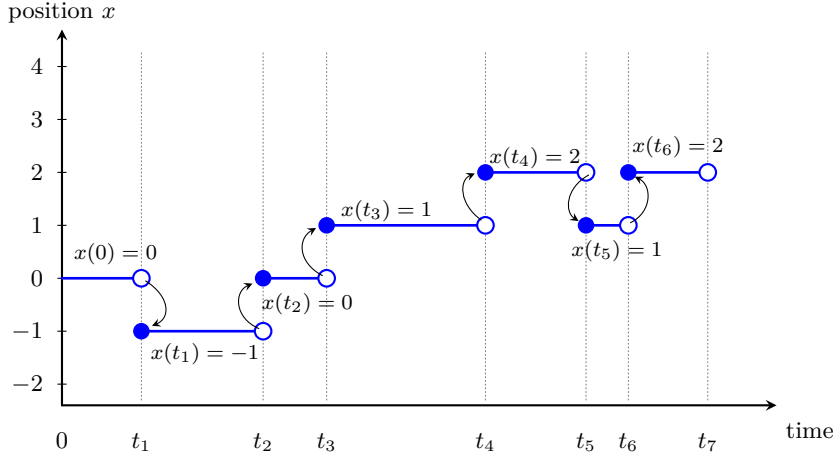

Consider a trajectory $\omega=(x(s),\,s\in[0,t])$ with $N_+$ jumps to the right
($x\to x+1$) and $N_-$ jumps to the left. Let
$N=N_++N_-$ be the total number of jumps (the activity) and $J=N_+-N_-$ the
time-integrated current. We ignore the initial condition, which is uniform in
the steady state. The weight of $\omega$ is the product of the rates of the
jumps that occur, times the probability $e^{-\xi t}$ of not jumping in between:
\begin{equation}\label{eq:path}
  \mathrm{Prob}[\omega]\propto p^{N_+}q^{N_-}\,e^{-\xi t}
  =: e^{-\mathcal{A}(\omega)} 
\end{equation}
With $N_\pm=(N\pm J)/2$ the action becomes
\begin{equation}\label{eq:action}
  \mathcal{A}=-\frac{J}{2}\log\frac{p}{q}-\frac{N}{2}\log(pq)+\xi t 
\end{equation}
The last term matters as soon as $\xi$ depends on $\mathcal{E}$ (it is
absent, or constant, only when it does not).\\

Time reversal $(\theta\omega)(s)=x(t-s)$ leaves $N$ and $t$ unchanged and flips
the sign of $J$. We split the action accordingly,
\begin{equation}\label{eq:split}
  \mathcal{A}=\mathcal{D}-\tfrac12\Sigma,\qquad
  \mathcal{D}=\tfrac12\bigl[\mathcal{A}(\theta\omega)+\mathcal{A}(\omega)\bigr],\qquad
  \Sigma=\mathcal{A}(\theta\omega)-\mathcal{A}(\omega)
\end{equation}
and read off from \eqref{eq:action}
\begin{equation}\label{eq:SigmaD}
  \Sigma(\omega)=\log\frac{\mathrm{Prob}[\omega]}{\mathrm{Prob}[\theta\omega]}
  =\beta\mathcal{E}d\,J,
  \qquad
  \mathcal{D}(\omega)=a\,N+\xi\,t,
  \qquad
  a:=-\tfrac12\log(pq)=\log\frac{2\cosh(\beta\mathcal{E}d/2)}{\xi}
\end{equation}
Hence
\begin{equation}\label{eq:A}
  \mathcal{A}=-\tfrac12\beta\mathcal{E}d\,J+a\,N+\xi\,t 
\end{equation}
The antisymmetric part $\Sigma$ is the heat dissipated into the bath divided by
$k_BT$, that is, the entropy flux per $k_B$. This is the local version of
microscopic reversibility mentioned above: mechanical time-reversal symmetry
leaving its trace on the mesoscopic dynamics~\cite{ldb,maes2002timereversalentropy}. If there is no driving, $\Sigma=0$. More generally, when the
walker moves in a potential $V$, one has
$\Sigma=\beta[V(x(0))-V(x(t))]$, a pure boundary term: this is detailed balance,
the exceptional situation in which the antisymmetric part carries no
information about the dynamics.\\
The symmetric part $\mathcal{D}$ is the frenesy. It depends on $N$ and on the
total time, but not on the direction of the jumps: it counts how much happens, not
which way. \\

Change the driving from $\mathcal{E}_0$ to $\mathcal{E}_0+\varepsilon$. To first order in $\varepsilon$ the
ratio of path weights is
\begin{equation}\label{eq:ratio}
  \frac{\mathrm{Prob}_{\mathcal{E}_0+\varepsilon}[\omega]}{\mathrm{Prob}_{\mathcal{E}_0}[\omega]}
  =1+\varepsilon\Bigl(\tfrac12\beta d\,J-a'N-\xi' t\Bigr)+O(\varepsilon^2)
\end{equation}
where primes denote derivatives with respect to $\mathcal{E}$ at $\mathcal{E}_0$. The
bracket has zero mean in the unperturbed steady state, as it should for a
normalized ratio: with $\langle J\rangle=(p-q)t$, $\langle N\rangle=\xi t$ and
$a'\xi+\xi'=\tfrac12\beta d\,(p-q)$ the three terms cancel. Writing
$\delta\Sigma=\varepsilon\,\beta d\,J$ and $\delta\mathcal{D}=\varepsilon(a'N+\xi't)$, the response of any
path observable $O$ is
\begin{equation}\label{eq:general}
  \langle O\rangle_{\mathcal{E}_0+\varepsilon}-\langle O\rangle_{\mathcal{E}_0}
  =\tfrac12\langle O;\delta\Sigma\rangle-\langle O;\delta\mathcal{D}\rangle 
\end{equation}
with the covariance $\langle A;B\rangle=\langle AB\rangle-\langle A\rangle\langle B\rangle$
in the unperturbed ensemble. This is the structure of Fig.~\ref{fig:dynfluct-scheme}: a dissipation excess
and a frenesy excess. For the current $O=J$ the term $\xi't$ drops out, because
$t$ is not random, and
\begin{equation}\label{eq:response}
  \langle J\rangle_{\mathcal{E}_0+\varepsilon}-\langle J\rangle_{\mathcal{E}_0}
  =\varepsilon\,\Bigl[\tfrac12\beta d\,\langle J;J\rangle-a'\,\langle J;N\rangle\Bigr]_{\mathcal{E}_0}
  +O(\varepsilon^2)
\end{equation}
The first (entropic) term comes from the time-antisymmetric part of the action,
the second (frenetic) one from the time-symmetric part.

Here everything can be computed. $N_+$ and $N_-$ are independent Poisson
variables with means $pt$ and $qt$, so
\begin{equation}
  \langle J;J\rangle=\xi\,t,\qquad
  \langle J;N\rangle=(p-q)\,t=\xi\,t\tanh\frac{\beta\mathcal{E}_0 d}{2}
\end{equation}
Inserting these in \eqref{eq:response}, with $a'=\tfrac12\beta d\tanh(\beta\mathcal{E}d/2)-\xi'/\xi$, gives the differential response per unit time,
\begin{equation}\label{eq:diffresp}
  \frac{\mathrm{d}}{\mathrm{d}\mathcal{E}}\frac{\langle J\rangle}{t}
  =\frac{\beta d}{2}\,\xi\,\mathrm{sech}^2\!\frac{\beta\mathcal{E}d}{2}
   +\xi'\tanh\frac{\beta\mathcal{E}d}{2}
\end{equation}
which agrees with differentiating $\langle J\rangle/t=\xi\tanh(\beta\mathcal{E}d/2)$ directly.\\

Near equilibrium, $\mathcal{E}_0=0$, we have $p=q$, so $\langle J;N\rangle=0$ (the
average of the time-antisymmetric variable $JN$ vanishes), and $a$ is even in $\mathcal{E}$,
so $a'(0)=0$. The frenetic term drops out and what remains is the
Sutherland--Einstein relation: with velocity $v=d\,J/t$ and diffusion constant
$D_0=\xi(0)d^2/2$, one finds $\mathrm{d}\langle v\rangle/\mathrm{d}\mathcal{E}=\beta D_0$, the mobility given by the
diffusion.

Away from equilibrium that is no longer true. The first term of
\eqref{eq:diffresp} is always positive, but $\xi'\tanh(\beta\mathcal{E}d/2)$ can be negative
and, if the escape rate drops fast enough with the driving, outweigh it. For
$\mathcal{E}>0$ the differential conductivity is negative as soon as
\begin{equation}
  \xi'<-\frac{\beta d}{2}\,\xi\,\frac{\mathrm{sech}^2(\beta\mathcal{E}d/2)}{\tanh(\beta\mathcal{E}d/2)} 
\end{equation}
In terms of \eqref{eq:response}: current and activity are positively correlated,
$\langle J;N\rangle>0$, and when $a'>0$ is large enough the frenetic term
$-a'\langle J;N\rangle$ is negative and wins~\cite{negative2002,neg}. The
entropic term alone could never produce this. The dissipation is the same in
both cases, only the way the system is kept busy differs, and that is a
frenetic effect. (Note also that, because $p/q$ is fixed by \eqref{eq:ldb},
holding $\xi$ fixed or holding the symmetric prefactor $\sqrt{pq}$ fixed are
different kinetic modelling choices, with different responses.)
\end{example}

Note that the analysis of the little model above uses trajectories and functions (like $N$ and $J$) on them.  Nonequilibrium analysis indeed deals with time, with the movie and not only with the single-time picture.  
A useful metaphor is a city at night.
Entropy is related to traffic imbalance: how many cars move from one district to another, and whether there is a preferred overall flow.  Frenesy is different. It measures how busy the city is overall: take the cinematic image of engines starting, brakes flashing, taxis turning corners, bicycles weaving between pedestrians. Even if the net traffic flow is zero, the city may still be alive with motion. Now imagine a morning when one of the main bridges is suddenly closed. In a city with little activity, the disruption remains local because there are few cars on the road. In a bustling city, drivers immediately search for alternative routes, congestion propagates rapidly, and traffic patterns reorganize throughout the network. The response depends not only on the net traffic flow but also on the underlying level of activity. Likewise, the response of a nonequilibrium system depends not only on entropy production but also on its frenesy.

\subsection{Time reversal {\it versus} field reversal}
The wake left behind a moving boat is a familiar image: retracing the
trajectory does not undo the eddies already created.
It useful to continue the previous random walk Example~\ref{rawa} for understanding the difference between field-reversal and time-reversal:\\
\begin{example}[Random walk, continued]\label{rawa2}
We compare two operations on the random walk: reversing time,
$\omega\mapsto\theta\omega$, and reversing the driving, $\mathcal{E}\mapsto-\mathcal{E}$.
Take the action
\begin{equation}\label{eq:Afull}
  \mathcal{A}_{\mathcal{E}}(\omega)=-\tfrac12\beta\mathcal{E}d\,J(\omega)+a(\mathcal{E})\,N(\omega)+\xi(\mathcal{E})\,t,
  \qquad a(\mathcal{E})=\log\frac{2\cosh(\beta\mathcal{E}d/2)}{\xi(\mathcal{E})}
\end{equation}
where we neglect the initial condition, which does not affect
the extensive part of the action. Since $J(\theta\omega)=-J(\omega)$ while
$N$ and $t$ are unchanged,
\begin{align}
  \mathcal{A}_{\mathcal{E}}(\theta\omega)&=+\tfrac12\beta\mathcal{E}d\,J+a(\mathcal{E})N+\xi(\mathcal{E})t,\\
  \mathcal{A}_{-\mathcal{E}}(\omega)&=+\tfrac12\beta\mathcal{E}d\,J+a(-\mathcal{E})N+\xi(-\mathcal{E})t 
\end{align}
The entropic terms agree. Only the frenetic (time-symmetric) terms are
different, so
\begin{equation}\label{eq:exactratio}
  \frac{\mathrm{Prob}_{-\mathcal{E}}[\omega]}{\mathrm{Prob}_{\mathcal{E}}[\theta\omega]}
  =\exp\Bigl\{\bigl[a(\mathcal{E})-a(-\mathcal{E})\bigr]N+\bigl[\xi(\mathcal{E})-\xi(-\mathcal{E})\bigr]t\Bigr\}
\end{equation}
That is exact. Both brackets are odd in $\mathcal{E}$, so the exponent starts at
first order and the next correction is $O(\mathcal{E}^3)$. Because $\cosh$ is even,
$a'(0)=-\xi'(0)/\xi(0)$, and with $\xi_0:=\xi(0)$ and $\langle N\rangle_0=\xi_0 t$,
\begin{equation}\label{eq:lin}
  \frac{\mathrm{Prob}_{-\mathcal{E}}[\omega]}{\mathrm{Prob}_{\mathcal{E}}[\theta\omega]}
  =\exp\Bigl\{2\mathcal{E}\,\xi'(0)\Bigl(t-\frac{N(\omega)}{\xi_0}\Bigr)+O(\mathcal{E}^3)\Bigr\}
\end{equation}
The exponent has zero mean, as it must, since both sides are normalized.\\

Here some consequences:
\begin{enumerate}
\item If $\xi'(0)=0$, then time reversal and field reversal are equivalent to
linear order, and in fact up to $O(\mathcal{E}^3)$. This holds in particular
when the medium is reflection symmetric, $\xi(-\mathcal{E})=\xi(\mathcal{E})$,
and then $p(-\mathcal{E})=q(\mathcal{E})$ gives
$\mathrm{Prob}_{-\mathcal{E}}[\omega]=\mathrm{Prob}_{\mathcal{E}}[\theta\omega]$ exactly,
for any nonzero $a$.
\item In general the two operations differ at linear order by the
\emph{frenetic} first-order term $2\mathcal{E}\xi'(0)\bigl(t-N/\xi_0\bigr)$, that is, by the part of
$\partial_{\mathcal{E}}\mathcal{A}$ that is even under $\theta$.
\end{enumerate}

We can be more explicit with a kinetic asymmetry.\\
Let the saddle point of the barrier between neighboring cells sit at a
fraction $\vartheta\neq\tfrac12$ of the bond (the barrier is not midway), so
\begin{equation}
  p=k\,e^{\beta\vartheta\mathcal{E}d},\qquad q=k\,e^{-\beta(1-\vartheta)\mathcal{E}d}
\end{equation}
which still satisfies local detailed balance, $\log(p/q)=\beta\mathcal{E}d$. Then
\begin{equation}
  \xi(\mathcal{E})=k\bigl(e^{\beta\vartheta\mathcal{E}d}+e^{-\beta(1-\vartheta)\mathcal{E}d}\bigr),\qquad
  \xi(0)=2k,\qquad \xi'(0)=k\beta d\,(2\vartheta-1)\neq0 
\end{equation}
The mean activity responds at first order,
\begin{equation}
  \frac{\langle N\rangle_{\mathcal{E}}}{t}=\xi(\mathcal{E})\simeq 2k\Bigl[1+(2\vartheta-1)\frac{\beta d\,\mathcal{E}}{2}\Bigr]
\end{equation}
The activity is invariant under time reversal, but the field-reversed
experiment has a different one, namely $\xi(-\mathcal{E})$. At linear order the two
differ by exactly the frenetic term in \eqref{eq:lin}. The
entropic part of the response is the same in both, since
$\langle N;J\rangle_0=0$: linear response of a time-symmetric observable
is purely frenetic.\\
Similarly, for spatially varying kinetics, the difference becomes very clear.\\
Let the bond $(x,x+1)$ carry the symmetric prefactor $\kappa_x$, so that the
rate across it is $\kappa_x e^{\pm\beta\mathcal{E}d/2}$ and the escape rate from $x$ is
$\xi_x(\mathcal{E})=\kappa_x e^{\beta\mathcal{E}d/2}+\kappa_{x-1}e^{-\beta\mathcal{E}d/2}$.
Time reversal leaves $\int_0^t\xi_{x(s)}\,\mathrm{d}s$ invariant and gives
$\mathrm{Prob}_{\mathcal{E}}[\theta\omega]/\mathrm{Prob}_{\mathcal{E}}[\omega]=e^{-\beta\mathcal{E}d\,J}$, while
\begin{equation}
  \frac{\mathrm{Prob}_{-\mathcal{E}}[\omega]}{\mathrm{Prob}_{\mathcal{E}}[\theta\omega]}
  =\exp\Bigl\{2\sinh\tfrac{\beta\mathcal{E}d}{2}\;\Phi(\omega)\Bigr\},
  \qquad \Phi=\int_0^t\bigl(\kappa_{x(s)}-\kappa_{x(s)-1}\bigr)\,\mathrm{d}s 
\end{equation}
which is again first order in $\mathcal{E}$ unless $\kappa_x\equiv\kappa$.
\end{example}
These effects are variations on the simpler truth that  time reversal
cannot change the stationary occupation densities, while field reversal may change it at linear order.

\subsection{Frenometry}
A natural question is whether --- and how --- frenesy can be measured. Measuring entropy and its production is, in general, also not direct: one measures heat capacity, or work and heat (e.g. via a Peltier element), and from these, changes in entropy are inferred. Something similar holds for frenesy. As seen in the example above, frenesy shows up in response, and frenometry may therefore isolate its contribution --- changes in frenesy, and how they correlate with observables.\\

Viscosity offers a natural entry point. It is, more generally, understood as the response of a time-symmetric observable, the momentum flux. Because the observable is time-symmetric, its response already picks up a frenetic contribution in the dynamical fluctuations, even around equilibrium.\\
Some physical effects, in fact, are fundamentally about dynamical activity --- most strikingly, how that activity collapses as viscosity increases without bound. The glass transition is the paradigm case: viscosity increases by some 15 orders of magnitude within a temperature window of some hundred kelvin. It is precisely the frequency of dynamical events, not a free-energy landscape alone, that controls the relevant timescale. While the jury is still out, we believe the glass transition fits within a purely kinetic (dynamical) theory, related to models of facilitated dynamics and kinetically constrained models, that is essentially frenetic: no underlying thermodynamic singularity is needed. Frenesy is then the real driver of the dramatic macroscopic phenomenon witnessed on cooling a glass-forming liquid.\\

There are also more immediate and almost tangible methods to measure frenesy.
For about a decade now, single-particle or single-molecule tracking has probably been the most direct route: in colloidal suspensions, active matter, and molecular motors, one can record trajectories and literally count the number of jumps, reversals, or configurational changes per unit time. That is essentially a direct measurement of dynamical activity, distinct from measuring net displacement, which reflects only the entropic or current part. Techniques where you get the rate of configurational change directly from the photon trace, independent of whether there is any net current or displacement, super-resolution tracking, and optical or magnetic tweezers already give dwelling times and escape-rate statistics for molecular motors, enzyme turnover, and ion-channel gating. Obviously, more recent experimental findings and technology such as the attosecond light pulses (Nobel Prize in Physics, 2023), optical traps (Nobel Prize in Physics, 2018) and the broader use of fluorescent dyes and lasers to track trajectories significantly add to our knowledge about dynamical activity.\\

In many-body systems, a related but distinct signature is accessible: how many particles are mobile at a given moment, and how that mobile population correlates in space and time, the subject of dynamic heterogeneity measurements in glasses. A different, and in some ways cleaner, instance of the same idea appears in counting statistics in mesoscopic and quantum transport, where one measures the rate and irregularity of discrete events crossing a junction.\\

Finally, on the proposal side, one could imagine designed ``frenetic probes'': small time-symmetric perturbations (e.g., periodic modulation of a trap stiffness or a barrier height, rather than a tilt or force) whose response isolates the frenetic term cleanly, analogous to how a tilt isolates mobility. As far as we know, that has not been systematically exploited experimentally, but it follows directly from the time-symmetry argument in the viscosity discussion above.  It is not purely speculative: e.g., negative (linear) friction on  a probe has already been predicted theoretically as resulting entirely from the frenetic contribution in the response of the nonequilibrium medium to the probe motion   \cite{Pei2025InducedFriction,Beyen2026Rayleigh}.

\subsection{Vis viva}
The intuition that ``how much is actually moving'' deserves its own conceptual category, distinct from static configuration, is not new. In the vis viva controversy of the late 17th and 18th centuries, Gottfried Leibniz (1686) distinguished {\it vis viva} (``living force,"  $\propto mv^2$, the ancestor of kinetic energy) from {\it vis mortua} (``dead force,'' the latent potential for motion contained in a body's position). The echo is suggestive: {\it vis mortua} is configurational and anticipates the static character of the entropic term, fundamentally about the shape of the landscape (Boltzmann entropy as a state count); {\it vis viva}, representing actual motion, anticipates frenesy, concerned with how much dynamical activity is occurring, independent of net direction.\\

The parallel is historical and thematic rather than mathematical. Frenesy is a property of trajectories, not an instantaneous kinetic quantity, but the impulse to separate the ``living'' (active) from the ``dead'' (passive) contribution to a system's behavior long predates dynamical fluctuation theory. Yet it resurfaces, in a more literal form, in the very language used today for large deviations of dynamical fluctuations: the corresponding rate functional is built action-like.  The analogy then is that the entropic term, fluxes times thermodynamic forces is the analogue of a magnetic field, breaking time-reversal symmetry.  The time-symmetric frenesy can be split in two components, a kinetic-activity term which is an even function of the fluxes and a potential-like term which takes more details on the reactivities and which we call the dissipation function/channel.  As a remarkable aside, the adiabatic invariance of entropy can be viewed as a consequence of Noether's theorem for a thermodynamic action \cite{Sasa_original,minami2020thermodynamic,beyen2025entropy,beyen2025noether_GENERIC}. \\
Nonequilibrium thermodynamics weighs the relative probability of many possible paths, with irreversibility built into that weighting from the start.  The zero-cost trajectory of the action (where the nonequilibrium Lagrangian  actually vanishes) delivers the macroscopic equation describing the evolution toward the most typical nonequilibrium condition \cite{M,maes2026relaxation,rao2016}.   

\section{A newer view on dissipation}\label{newe}
Nature’s default trajectory is toward higher entropy, and living nature is fighting it, so it could seem:
\begin{quote}
The general struggle for existence of animate beings is not a struggle for raw materials --- these, for organisms, are air, water and soil, all abundantly available --- nor for energy, which exists in plenty in any body in the form of heat, but a struggle for [low] entropy, which becomes available through the transition of energy from the hot Sun to the cold Earth.\\
{\it The quote originates from Boltzmann's 1886 lecture
``Der zweite Hauptsatz der mechanischen Wärmetheorie'',
later translated in Ref.~\cite{Boltzmann1886}}.
\end{quote}
Indeed, in many discussions over the last centuries, the increase of entropy has been one of science’s most unsettling truths\footnote{As an extreme example of mental disarray, some young-Earth creationist writers have argued that the physical ``curse'' placed on creation after the Fall (the original sin according to christian theology) brought the Second Law into effect, transforming a static or perfect universe into one subject to decay.}. As early as 1851, Lord Kelvin (William Thomson) described the ultimate fate of the universe as a heat death\footnote{A heat death is a state of maximum entropy given fixed total energy, in which cooling radiation is the visible tracer.}``a universe ending in permanent equilibrium.''  The story actually began earlier with a young French engineer named Sadi Carnot. In 1824, while analyzing heat engines, Carnot recognized that no machine could be perfectly efficient, even when running close to zero power. There is always a cost; energy is inevitably lost to friction and sinks. Decades later, Clausius formalized this concept as entropy, concluding that the total entropy of the universe is constantly expanding toward a maximum. In that way, for much of the history of thermodynamics, dissipation was regarded primarily as a process of energy degradation and the loss of mechanical organization.  And for a long time, entropy was synonymous with messiness, unrecoverable losses, and decay.\\

Today the emphasis has changed. Rather than viewing dissipation merely as a source of loss, we increasingly recognize it as the very condition that makes organized nonequilibrium behavior possible. Indeed, Carnot had already recognized in his abstract study of heat engines that the low-temperature bath is not merely a heat sink for the unavoidable degradation of energy but an essential ingredient for work to be produced at all. Motion in a steam engine requires not only a hot source but also a cold sink. When saying that a heat engine can never be free of losses, one may add that, when properly balanced, those very losses are what make the engine possible in the first place.

Nearly one hundred and fifty years after these {\it Réflexions sur la puissance motrice du feu}, the Nobel Prize in Physiology or Medicine was awarded to the biochemist Christian de Duve for the discovery of lysosomes, the microscopic recycling and waste-processing centers of the cell. These intracellular organelles act as the cell's digestive and sanitation system, continuously degrading and recycling cellular material, and are indispensable for cellular survival.  Biological life, too, depends not only on construction but equally on controlled degradation.\\

However, there is more to be said, and that is taken up in the following subsections.  The exciting point is that interesting new behavior may result in dissipative systems. That requires much more analysis, and progress is to be expected first for more restricted classes of nonequilibrium systems.  In a way, there is far too much of nonequilibrium...  The phenomenology is so terribly diverse and the kinetics so broad, that it is hard to imagine a powerful framework.  Nevertheless, people have tried, and in various directions. Nowadays, it appears fair to say that systems that are characterized by local detailed balance in their weak coupling with spacetime well-separated reservoirs, make a class apart suitable for a unifying description.\\

Local detailed balance appears compatible with the class of behavior as witnessed in so-called `dissipative structures,' obtained as a new regime after a dissipation-induced instability or as driving-induced stabilization of what was forbidden under equilibrium.  That has drawn sustained interest for roughly seventy years.  It is important because, without mechanisms of self-organization driven by nonequilibrium dynamics, the emergence of persistent order in nature becomes considerably more mysterious.  Ignoring such a view, disabling the role of self-organization and nonequilibrium in general, may lead people to turn to anthropic principles or to intelligent design to understand the prevalence of ordered structure and the origin or evolution of life.

\subsection{Dissipation-induced instability, requiring driving}
The qualitative role of dissipation had already been appreciated much earlier. In the first edition of the ``Treatise on Natural Philosophy'' (1867) \cite{ThomsonTait1867}, Thomson and Tait emphasized that friction fundamentally changes mechanical evolution by converting organized motion into heat and thereby selecting particular dynamical behaviors.\\
Toward the end of the nineteenth century, another important insight emerged with the discovery of instabilities associated with and even driven by heat flow. 
A dissipation-induced instability is the counterintuitive phenomenon where a system that is stable (or marginally stable, e.g. via a conserved Hamiltonian structure) with a given amount and type of dissipation becomes unstable once a small amount of (other type of) dissipation is added, like a spinning satellite with internal energy dissipation destabilizing an otherwise stable rotation about the minor axis\footnote{This was discovered, dramatically and unexpectedly, with Explorer 1 (the first US satellite, 1958): designed to spin about its long, minor-inertia axis (a thin cylinder, spin-stabilized like a bullet), it instead began tumbling end-over-end within hours after launching, settling into rotation about its major axis. That is similar to the tippe top. The culprit was flexing of its four flexible whip antennas, which dissipated energy internally.} \cite{Krechetnikov2007}. There, dissipation is the destabilizing agent, against naive intuition that removing energy should always calm things down. \\

Things become more interesting when there is driving.  Then, these dissipation-induced instabilities are the thermodynamic counterpart of what is known as the Ziegler paradox \cite{Ziegler1952}: a driven system that is perfectly stable  may become unstable when an arbitrarily small amount of (some type of) damping is introduced.  The point to understand here is that a given amount of driving does not produce a fixed dissipation.  How much heat is produced and how the distribution of dissipation is organized depends on time-symmetric aspects. Then, in nonequilibrium studies, we call an instability dissipation-induced when at fixed (maybe high-enough but perhaps also only initial) driving a small amount of dissipation makes the system unstable \cite{Soriente2021dissipation,Krechetnikov2007}. The stability boundary responds discontinuously, often non-monotonically, in the dissipation parameters.  We repeat (and emphasize)  that, for driven systems, it is the distribution of damping across degrees of freedom, not only its magnitude, that decides the outcome. It is for instance the structure of the dissipation matrix relative to the driving, and not its amount that causes the instability.  That is the reason why we add the ``frenetic selection'' on top of Fig.~\ref{fig:frenetic-scheme}.
 \begin{figure}[htbp]
\centering
\includegraphics[width=0.6\textwidth]{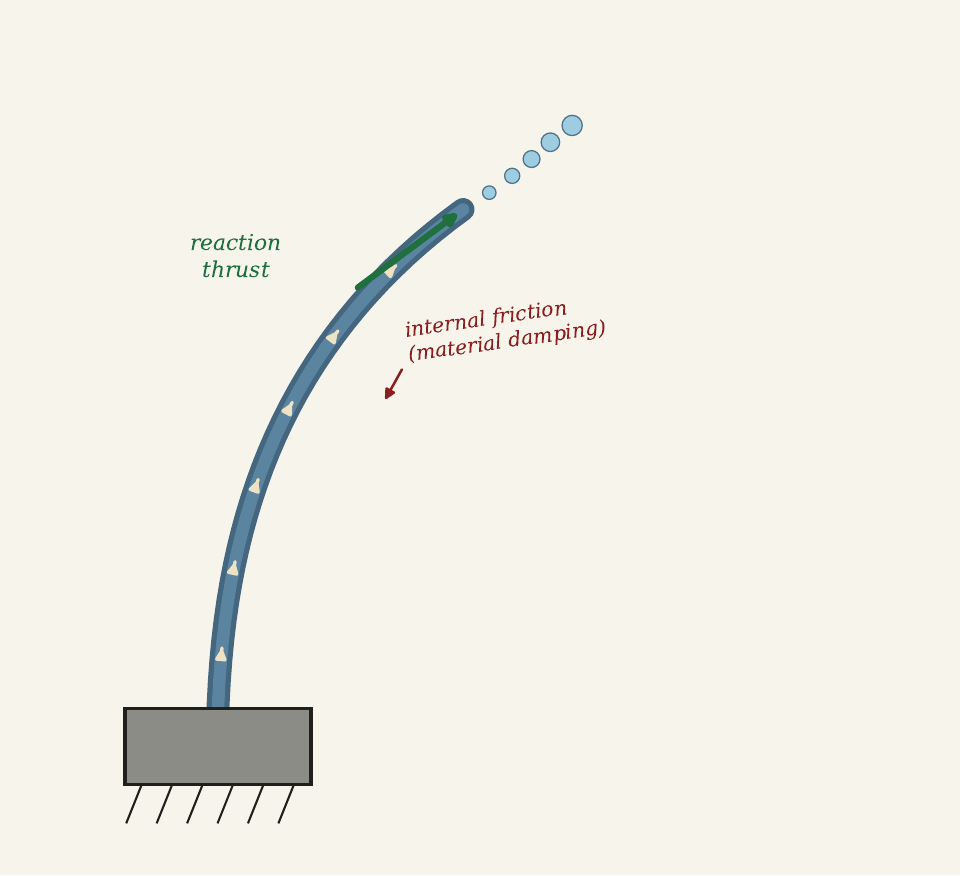}
\captionsetup{justification=raggedright, singlelinecheck=false}
\caption{\small{Ziegler's column realized as a garden hose with a free end. Above a critical flow rate, the straight configuration becomes dynamically unstable and the hose whips. A stiff nozzle attached to an otherwise floppy hose, or a patch of stiffer material partway along its length will create more flutter. The water flowing through the hose carries kinetic energy.  As the tip of the hose curves, the water exits at an angle. This exerts a reaction force (thrust) that always remains tangent to the tip of the hose. At a critical flow velocity, the energy added to the hose by this nonconservative follower force exceeds the natural structural damping of the rubber. The system enters a state of effective ``negative damping,'' causing the whipping motion to grow violently.}}\label{ho}
\end{figure}
 
\begin{example}[Ziegler column: the garden hose]
See Fig.~\ref{ho}. The damping coefficients themselves at the joints, or the material/fluid damping along the hose stay strictly positive. The instantaneous dissipation rate is always nonnegative; no hidden negative-friction element anywhere.  What is important is that the follower force $P$ is nonconservative; we are dealing with a nonequilibrium system and  work over a cycle of oscillation is path-dependent. The total energy balance
$$\frac{\id}{\id t}(\text{kinetic} + \text{potential}) = W_P - D$$
contains the work $W_P$ done by the follower force $P$ and $D\ge 0$ is the dissipation. $W_P$ over one oscillation period need not vanish and crucially, how much net positive work $P$ manages to extract from the motion depends on the phase relationship between  the local bending angles along the hose. Adding a small amount of damping shifts that phase relationship, and it can shift it in exactly the direction that lets $P$ pump in more energy than the dampers remove: $W_P > D$, even though $D>0$ throughout. It is the  ``pattern of damping'' rather than its magnitude but enabled by the damping, the energy exchange gets reshaped between modes to create the effective negative damping.  That shape-dependence, rather than magnitude-dependence, of damping in Ziegler's paradox is reminiscent of and perhaps even a deterministic-mechanical echo of the way frenesy captures pattern-sensitive, time-symmetric aspects of dynamics that entropy production alone cannot see.\\
It may appear strange to speak about frenetic selection for Ziegler's column, as it is standardly treated in terms of a deterministic linear ordinary differential equation.  There is nothing stochastic, some would say.  Yet, part of the main thesis is that this evolution carries the frenetic signature simply because it gives the typical evolution among many other possible trajectories with a distribution (of the dynamical fluctuations) that is governed by both (trajectory-dependent) entropy production and frenesy \cite{maes2026relaxation}.  Of course, that claim needs an explicit calculation showing the destabilization threshold is recoverable from, or naturally expressed in terms of, the frenetic content of some specified noisy completion.
\end{example}

Such instabilities (as in the Ziegler example above and in Fig.~\ref{ho}) need not be viewed as destructive or catastrophic. On the contrary, they often herald the emergence of qualitatively new behavior in the nonlinear saturated regime. A stationary mechanical system may spontaneously develop organized oscillations, while a steady flow can become the source of persistent currents and self-sustained activity. Instability thus came to be understood not as the failure of order, but as its birthplace.

The discovery and understanding of dissipation-induced instabilities changed into a recognition of the constructive role of maintained dissipation; it may allow surprising and quite fascinating behavior. 

\begin{example}[Tippe top]
Dissipation is not merely the enemy of organization but frequently its architect, much as the celebrated tippe top which, with only the slightest contact friction, overturns to spin gracefully on its narrow stem, an effect first patented by Helene Sperl in 1891. Yet the tippe top's flip is a one-shot relaxation, powered by a single finite reservoir, the initial spin, not a continuously maintained current. Once it flips and slows, it simply stops; strictly, it does not qualify as a dissipative structure. Setting that difference with Ziegler's column aside, the existing explanation of the tippe top's instability is complete, rigorous, and entirely classical: a conserved Jellett quantity, a monotonically decreasing energy, and nothing more. No path measure, no action functional, no appeal to frenetic control is needed anywhere.\\
But this classical completeness should not obscure a deeper point: such a macroscopic (deterministic) trajectory can be understood as the typical path singled out, via a law-of-large-numbers argument, from an ensemble of fluctuating paths whose likelihood is governed jointly by entropic and frenetic content\footnote{It is the deeper ground of fluctuation-response relations, whether in or out of equilibrium.  Macroscopic response and evolution is decided by the ingredients of the dynamical fluctuation functional.}. Nothing in Ziegler's column, the tippe top, or Explorer 1 strictly requires this vocabulary. Each has its own self-sufficient, mechanical explanation. What the idea of frenetic selection offers instead is a unifying lens on what a dissipation-induced instability really is.\\
Then, for the tippe top, what matters is not dissipation as such but its location. Air drag also removes energy.  It is dissipative in exactly the thermodynamic sense.  Yet, it does not cause the inversion, because it acts nowhere near the torque that couples to the tipping angle. It is the contact friction specifically, positioned exactly where the constraint and the destabilizing torque coincide, that drives the instability. We claim this locational selectivity is itself a signature of the frenetic content of the underlying dynamical fluctuations.  That is not offered as a new explanation of the tippe top, which needs none, but as a way of seeing, in one language, why dissipation-induced instabilities pick out the particular channel of dissipation that they do. 
\end{example}

There are of course other scenarios.  It may be that structure appears only for strong enough driving, subject of the next subsection.  The driving then stabilizes what was unstable without dissipation, such as a paramagnetic phase coexisting with ferromagnetic phases.  This will complete the double origin of dissipative structures as schematized in Fig.~\ref{fig:frenetic-scheme}: driving stabilizes what dissipation would ruin {\it versus} dissipation destabilizes what conservative dynamics would preserve.

 \subsection{Driving-induced stability, requiring dissipation}
 Over the last 70 years, in much stronger terms, time-reversal symmetry breaking and dissipation --- the production of entropy --- is being recognized as an important source (not the agent) or facilitator of interesting and useful functioning: in a world far from passive equilibrium, steady entropy production seems correlated with or somehow allows nature to spontaneously build complex, beautiful structures that could never exist otherwise.\\
 That often goes via instabilities (exactly as above) but a different emphasis brings us to an important class of systems where, for given dissipation channels, the driving amplitude matters to stabilize what is unstable under equilibrium. Early examples of such driving-induced instabilities include the thermo-acoustic instability studied by Lord Rayleigh \cite{Rayleigh1878Explanation,Rayleigh01041883} and the convective instability observed by Henri Bénard in 1900 and later explained by Rayleigh \cite{Benard1900,Rayleigh1916} in which the quiescent conductive state between a hot plate on the bottom and cold plate on top becomes linearly unstable\footnote{Bénard's 1900 cells were surface-tension driven (Marangoni); Rayleigh’s 1916 explanation is for buoyancy-driven convection.}.\\
 
 Such instabilities may ``open'' the stationary system to a new interesting phase or condition away
from equilibrium, called dissipative structures.  Such a view was entertained explicitly and predominantly by Nobel laureate
Ilya Prigogine and the Brussels school \cite{Prigogine1978Nobel,Nicolis1977SelfOrganization},
with closely related and equally influential routes opened, from physics, by Hermann
Haken's theory of synergetics \cite{Haken1975RMP,Haken1977Synergetics}, from
developmental biology by Alan Turing's theory of morphogenetic pattern formation
\cite{Turing1952}, and from prebiotic chemistry by Manfred Eigen's theory of
self-organizing molecular evolution \cite{Eigen1971Selforganization,Eigen1979Hypercycle}.
Concrete experimental realizations of such dissipative structures\footnote{Such structures (Bénard rolls, chemical patterns, sustained oscillations, active-matter patterns,...) are called dissipative (by Prigogine) because they decay when you turn off the driving.} followed, for instance
in the form of the oscillating Belousov--Zhabotinsky
reaction\footnote{Belousov's original work was rejected by journals in the 1950s and
only circulated informally before being posthumously published.} \cite{Belousov1985,Zhabotinsky1964}.
Ultimately, active matter, from bacteria to humans, requires dissipation to thrive.   As a matter of fact, the entire history of our universe is dissipative, with an early universe being very far from equilibrium.\\

It is useful here to emphasize the difference with Floquet engineering and (purely) driving-induced effective stabilization.  As an example of that mechanics where rapid periodic forcing gets ``rectified'' by nonlinearity into an effective, static-looking restoring force, we may mention the Kapitza pendulum.  It  is (near-)Hamiltonian: no friction, no noise, no heat bath, and crucially no entropy production at all. Instead, for thermodynamic purposes as above and for the terminology of dissipative structure,
we insist on stabilization or order that requires ongoing dissipation.

\subsection{Selection}

The difference between the mechanisms in the two previous sections is not very big (but subtle).  In the dissipation-induced case, we fix the driving (large enough) and we search for or simply turn on dissipative channels that make the (original) system unstable, after which a new interesting stationary condition emerges.  On the other hand, in the driving-induced case, we fix a dissipation channel, and we see how by increasing the driving new structures or phases appear.  In both cases but not always equally visible, the channel or the location or the kinetics of the dissipation mechanism matters a lot.  In other words, the very presence of entropy production is not determining; there is a selection mechanism. For instance, we may find that by driving a magnet (switching temperatures or time-dependent magnetic fields) there appears a genuine coexistence of paramagnetic and ferromagnetic phases (breaking the Gibbs phase rule of equilibrium thermodynamics).  However, the appearance of this coexistence may depend on kinetic parameters in the dynamics --- transition temperatures shift and phases come and go depending on time-symmetric reactivities. Referring to Fig.\ref{fig:frenetic-scheme}, we ask what paves the road?\\

Lifetimes and escape rates, how long a system dwells in a given configuration before leaving it, together with the accessibility of different configurations and the overall dynamical activity sustaining transitions between them, come to the foreground now. As written before, entropy production often tells us where processes prefer to go, while frenesy tells us how intensely the system explores possibilities in getting there.
Accordingly, nonequilibrium should not be understood primarily through a state that dissipates, but through the process itself: the change, the kinetic selection and activity along trajectories. For structures maintained by continuous dissipation, breaking time-reversal symmetry makes possible forms of spatial, internal, and chiral symmetry breaking that equilibrium thermodynamics either forbids outright or can produce only through the thermodynamic limit, $N\to\infty$. Nonequilibrium needs no such limit for the open system itself: a single driven particle, or a lone molecular motor, can already show breaking of time-reversal and parity symmetry, sustaining a directed current where no equilibrium analogue at any system size exists. We propose that frenesy is what selects among the competing broken-symmetry structures that this widened possibility space allows.\\

A flute provides a familiar metaphor. An unblown flute, unless enchanted\footnote{as the one of Prince Tamino.}, remains silent. But a steady stream of air, blown across the embouchure, excites a broadband, unstable jet oscillation, a dynamical activity with no melody of its own, exploring a whole range of frequencies rather than committing to one. It is the pipe's resonant boundary conditions, not the airflow, that select which of these frequencies survives as a sustained note. Opening or closing a hole changes the local escape rate of acoustic energy from the bore, shortening or lengthening the effective resonator and so retuning which mode gets selectively amplified. In this sense the flow enables the instrument's geometry to perform the selection. Dissipation drives the oscillation, but the escape gates, the narrow doors of Section \ref{bol},  choose the structure it settles into.\\

Likewise, breaking time-reversal symmetry in a driven chemical network as in the cyclic, ATP-consuming conformational changes of a molecular motor, can generate directed transport of matter coupled to that chemistry, with no equilibrium counterpart regardless of system size. Such mechanisms underlie chemo-mechanical coupling, mechanobiology, and many other nonequilibrium processes in which dissipation not only drives motion but also selects structure and function.\\
The biosphere of planet Earth originated in the flow of low-entropy energy coming from the Sun and high-entropy energy being radiated to cold interstellar space.  That dissipative process permits the time-symmetric fluctuation sector, frenesy, to select populations and evolutions\footnote{Selection of nonequilibrium evolutions may not happen at all, by the nonapplicability of the law of large numbers. For instance, turbulence in the inviscid limit is a case where the very mechanism that makes frenetic selection meaningful elsewhere, a unique dominant trajectory, breaks down, and that breakdown is the physically remarkable content of spontaneous stochasticity.}. We may speculate that this is what makes evolution and life possible as we know it. Perhaps one may think of life as matter that has learned to harness frenesy, and has come to optimize the combination of dissipation and activity parameters for specific structure and functioning.  
\begin{quote}
To determine whether under a given set of planetary conditions life is the preferred state or only a metastable state, we cannot just compare the lifeless state and the known biological state, but must consider the transitions between these states.\\  
{\it Charles Bennett, as quoted by Rolf Landauer in \cite{landauer}.}
\end{quote}
Indeed, the plausibility of life in this universe or on this planet should not be counted in terms of Boltzmann entropy, where we use a principle of indifference for all ways of combining material.\\

Finally, look back at Fig.~\ref{fig:toroidal}, and imagine that one of the connections, say between $x$ and $x+1$,  is somewhat congested, like in our house with narrow doorways of Section \ref{bol}.  Because of the bias, the walker will then spend a long time in cell $x$, and the stationary density of colloids gets a peak at $x$.  That is an example of the Blowtorch Theorem \cite{landauer,heatb,Landauer1993Machinery}:  a local kinetic bottleneck can pile up probability even when the equilibrium profile would be flat, precisely because breaking time-reversal symmetry makes the system sensitive to where the friction is, not just how much there is. Before, the distribution was uniform, and also for zero bias the stationary density profile does not change by the local congestion.  Only because of the $\cal E \neq 0$ (and the implied breaking of time-reversal symmetry) do we get a more interesting response and can we control the profile via frenetic steering \cite{LefebvreMaes2024}.\\

\section{Main thesis and open questions}\label{math}
In equilibrium physics, frenesy often hides in the background because detailed balance makes many processes appear calm and reversible. It does show up in relaxation behavior and for dynamical correlations. But in steady nonequilibrium systems, frenesy emerges as a central actor:
it captures the time-symmetric aspect of dynamics, complementing entropy production in determining the plausibility of trajectories. For a broad class of nonequilibrium phenomena, that perspective provides a more unified framework for describing the statistics of mesoscopic and macroscopic systems driven away from equilibrium. One of its most important consequences is that response to external perturbations is systematically governed jointly by entropy production and frenesy, thereby extending the familiar relation between fluctuations and response far beyond equilibrium.\\

Hence, interestingly, dissipation empowers an additional ingredient: the time-symmetric aspects of the dynamics. Beyond response, they allow the selection of population, current statistics and more general aspects by determining the specific dissipation channels, following the ideas of the older blowtorch theorem. Frenesy, as an umbrella term for time-symmetric contributions, thus enables a form of control or even steering.   Nature is thereby endowed with a formidable mechanism.  Remarkably, or even paradoxically, the breaking of time-reversal symmetry may foster the breaking of spatial and internal symmetries, thanks to time-symmetric ingredients.  We propose the latter indeed play a key role in determining which of the competing structures is ultimately selected. Low-entropy resources are precious, but it is frenesy that redistributes them among competing degrees of freedom. \\

Much is left to be understood (and to be calculated).  We mention two important open questions.\\  First, it would be useful to generalize notions of frenesy to the quantum world.  As mentioned before, time and trajectory are neglected in standard quantum mechanics, and dynamical fluctuation theory is largely inexistent. There are various proposals how to make sense of quantum trajectories\footnote{Characterizing trajectories via the Feynman-Vernon action \cite{Aurell2023QuantumFV,Strasberg2022QuantumStochastic} or via Bohmian mechanics \cite{DurrTeufel2009Bohmian} appears most promising.  At any rate, the fact that quantum processes have memory (ignored for instance in descriptions using Lindblad dynamics) does not help.  Surprisingly, there are exquisite analogies between the zig-zag dynamics of electrons and run-and-tumble processes; see for instance \cite{cartier06,penrose04,Krekels2024ZigZag,Maes2022RunAndTumble}.} which is necessary for studying dynamical activity at all.  We still hope for a breakthrough to add to and even to revolutionize the enterprise of steady-state quantum nonequilibrium physics and corresponding low-temperature phenomena for nonequilibrium photons, phonons and cold gases. We posit that trajectories of electrons and fermions more generally will be crucial there, as this will enable discussing their time-symmetric dynamical activity.\\ 

Secondly, the way to proceed with frenesy is operationally problematic when faced with many degrees of freedom.  We can always try to reduce the description to a smaller set of collective (but still fluctuating) variables, but at some moment interactions between many degrees of freedom must enter. It is not clear how to associate dynamical activity to a large collection of such particles in an experimentally or operationally meaningful way.  The kinetics becomes overwhelming, so to speak. Here, it may be simpler to take the ``infinite limit' immediately, and work with nonequilibrium field theories where the many-body aspect can be replaced with nonlinearities.  That ``replacement'' is already in itself a challenging project.  Yet, very promising work is done on steady state thermodynamics, including \cite{Sasa2006SteadyStateThermo,Nakagawa2019GlobalThermo,Nakamura2008FluctuationResponse}, and similarly, many results have appeared on active surfaces or membranes, including \cite{Salbreux2017ActiveSurfaces,AlIzzi2021ActiveMembranes,Mietke2019MinimalModel,Sahu2017Irreversible,AlIzziMorris2021Review}.  Those results are leading to question on how to organize a dynamical fluctuation theory for geometric degrees of freedom, and what then is the nature of geometric dynamical activity.  There, fluctuating space and the bustle of time will be meeting.

\bibliographystyle{unsrt}  
\bibliography{refs}

\end{document}